\documentclass[10pt]{article}
\RequirePackage{lineno}
\usepackage{amsmath}
\usepackage{amssymb}
\usepackage{hyperref}
	\newcommand{\rube}[1]{\emph{C.~rubella#1}}
	\newcommand{\grand}[1]{\emph{C.~grandiflora#1}}
	
	\newcommand{\figref}[2]{Figure \ref{#1}#2}
\usepackage{graphicx}
	
	\newcommand{\yb}[1]{{ \color{blue} #1}}

\usepackage{cite}
\usepackage[usenames,dvipsnames]{color}
\usepackage[usenames,dvipsnames]{xcolor}
\usepackage[labelfont=bf,labelsep=period,justification=raggedright]{caption}
\makeatletter
\renewcommand{\@biblabel}[1]{\quad#1.}
\makeatother
\setpagewiselinenumbers
\modulolinenumbers[5]
\linenumbers

\date{}
\begin{document}
% Title must be 150 characters or less
\begin{flushleft}
{\Large
\textbf{Genomic identification of founding haplotypes reveals the history of the selfing species \emph{Capsella rubella }}
}
\newline
% Insert Author names, affiliations and corresponding author email.

Yaniv Brandvain$^{1,\ast}$, 
Tanja Slotte$^{2,3}$, 
Khaled M. Hazzouri$^{4}$,
Stephen I. Wright$^{5}$,
Graham Coop$^{1}$
\newline 

\bf{1} Department of Evolution and Ecology \& Center for Population Biology, University of California - Davis, Davis, CA, USA
\\
\bf{2} Department of Evolutionary Biology, Evolutionary Biology Centre, Uppsala University, Uppsala, Sweden
\\
\bf{3} Science for Life Laboratory, Uppsala University, Uppsala, Sweden
\\
\bf{4} Department of Biology, New York University -- Abu Dhabi, Abu Dhabi, United Arab Emirates.
\\
\bf{5} Department of Ecology and Evolutionary Biology, University of Toronto, Toronto, ON, Canada
\\
$\ast$ E-mail: ybrandvain@gmail.com
\end{flushleft}

\newpage
%%%%%%%%%%%%%%%%%%%%%%%%%%%%%%%%
% ABSTRACT
% Please keep the abstract between 250 and 300 words
%%%%%%%%%%%%%%%%%%%%%%%%%%%%%%%%

\section*{Abstract}
The shift from outcrossing to self-fertilization is among the most common evolutionary transitions in flowering plants.
Until recently, however, a genome-wide view of this transition has been obscured by both a dearth of appropriate data and the lack of appropriate population genomic methods to interpret such data. 
Here, we present a novel population genomic analysis detailing the origin of the selfing species, {\em Capsella rubella}, which recently split from its outcrossing sister,  {\em Capsella grandiflora}. 
Due to the recency of the split, much of the variation within  {\em C. rubella} is also found within {\em C. grandiflora}.
We can therefore identify genomic regions where two {\em C. rubella} individuals have inherited the same or different  segments of ancestral diversity (i.e. founding haplotypes) present in {\em C. rubella}'s founder(s).  
Based on this analysis, we show that  {\em C. rubella} was founded by multiple individuals drawn from a diverse ancestral population closely related to extant  {\em C. grandiflora}, 
	that drift and selection have rapidly homogenized most of this ancestral variation since \emph{C. rubella}'s founding, 
	and that little novel variation has accumulated within this time. 
Despite the extensive loss of ancestral variation, the approximately 25\% of the genome for which two \emph{C. rubella} individuals have inherited different founding haplotypes makes up roughly 90\% of the genetic variation between them. 
To extend these findings, we develop a coalescent model that utilizes the inferred frequency of founding haplotypes and variation within
  founding haplotypes to estimate that {\em C. rubella} was founded by a
potentially large number of individuals between 50 and 100 kya, and
has subsequently experienced a twenty-fold reduction in its
effective population size. 
As population genomic data from an increasing number of outcrossing/selfing pairs are generated, analyses like the one developed here will facilitate a fine-scaled view of the evolutionary and demographic impact of the transition to self-fertilization.

%\newpage 

%%%%%%%%%%%%%%%%%%%%%%%%%%%%%%%%
% AUTHOR SUMMARY
%%%%%%%%%%%%%%%%%%%%%%%%%%%%%%%%

%\section*{Author Summary}
%While many plants require pollen from another individual to set seed, in some species self-pollination is the norm. 
%This evolutionary shift from outcrossing to self-fertilization is among the most common transitions in flowering plants. 
%Here, we use dense genome sequence data to identify where in the genome two individuals have inherited the same or different segments of ancestral diversity present in the founders of the selfing species, {\em Capsella rubella} to obtain a genome-wide view of this transition. 
%This identification of founding haplotypes allows us to partition mutations into those that occurred before and after {\em C. rubella} separated from its outcrossing progenitor, {\em C. grandiflora}.
%With this partitioning, we estimate that {\em C. rubella} split from {\em C. grandiflora} between 50 and 100 kya. 
%In this relatively short time frame, an extreme reduction in {\em C. rubella}'s population size is associated with a massive loss of genetic variation and an increase in the relative proportion of putatively deleterious polymorphisms. 

\newpage

%%%%%%%%%%%%%%%%%%%%%%%%%%%%%%%%
% INTRODUCTION
%%%%%%%%%%%%%%%%%%%%%%%%%%%%%%%%

\section*{Introduction} 
Most flowering plants are hermaphroditic, but many have evolved elaborate mechanisms to avoid self-fertilization and the associated costs of inbreeding \cite{Darwin1862,Darwin1876}. 
However, an estimated $15\%$ of flowering plant species are predominantly self-fertilizing  \cite{Goodwille2005, Igic2006} 
	and many of these species have evolved floral morphologies that promote this means of reproduction. 
This shift from outcrossing to inbreeding by self-fertilization is among the most common transitions in flowering plants \cite{Stebbins1950,Stebbins1974}, 
	and can occur when the short-term benefits of selfing (e.g. assured fertilization \cite{Baker1955}, 
	the `automatic' transmission advantage \cite{Fisher1941}, 
	and the maintenance of locally adapted genotypes \cite{Schoen1984}) 
	overwhelm the immediate costs of inbreeding depression \cite{Lande1985, Charlesworth2006}. 
However, in the longer term, limited genetic diversity and difficulty in shedding deleterious mutations 
	are thought to doom selfing lineages to extinction \cite{Stebbins1957,Takebayashi2001,Goldberg2010}.

While the causes and consequences of plant mating system evolution have long fascinated evolutionary biologists,
	the paucity of population genomic data for species with a recent shift in mating system and an absence of a framework 
	in which to interpret such data have prevented the development of a genome-wide understanding of this transition.
Here, we introduce a novel approach that utilizes patterns of variation in a recently derived selfing population to partition diversity within and among founding haplotypes. 
By partitioning two sources of sequence diversity -- incompletely sorted ancestral polymorphisms  and \emph{de novo} mutations which occurred since the population origin -- 
 	we generate a novel view of the selective and demographic history of a recently derived selfing population.  
In particular, we can  distinguish two factors that can lead to low
diversity in selfers: 
	the loss of ancestral polymorphism that occurred at the transition to selfing 
	and a long term small effective population size since the transition.

We apply this framework to the selfing species, {\em Capsella rubella},     
	for which we make use of a recently available population genomic dataset \cite{Wright2012}  
	%({\bf \color{red} Note to reviewers, this `genome paper', Slotte et al. 2013, is attached as a supplementary file})
	consisting of eleven resequenced transcriptomes -- six of {\em C. rubella} and five of a closely related, 
	obligately outcrossing species, \emph{C. grandiflora}, 
	to generate a well-resolved, genome-wide view of the
        transition from outcrossing to selfing and its immediate consequences.
While the origin of {\em C. rubella} has received significant attention \cite{Foxe2009,Guo2009,Wright2012,StOnge2011}, our understanding of {\em C. rubella}'s 
	history has been hampered by the small number of independent loci examined in previous studies and by the lack of methods tailored to understand the somewhat  
	unusual haplotype structure of genetic variation within recently derived selfing species.  
Similarly, while \emph{C. rubella} contains relatively elevated levels of putatively deleterious variation \cite{Foxe2009,Guo2009,Wright2012}; 
previous analyses could not partition  the extent to
which this was due to a long-term relaxation of the efficacy of purifying selection, or extreme sampling variance at the founding of the species.
Perhaps the most intriguing `origin story' for \emph{C. rubella} argues that at the last glacial maxima,  
	a single individual capable of selfing may have became isolated and gave rise to the entire species \cite{Guo2009}.  
Evidence for this hypothesis comes from the observation of only one or two distinct haplotypes per a locus in a sample of 17 loci examined in 25 \emph{C. rubella}  individuals  \cite{Guo2009}.

%Enabled by the novel descriptive statistics generated by our method, 
Here, we use our novel framework and coalescent modeling to investigate the origin of \emph{C. rubella}  focusing on: 
	testing the hypothesis that it was founded by a single individual, 
	estimating  the timing of its founding, 
	comparing patterns of variation across its distribution, 
	estimating its long-term effective population size, 
	and documenting the weakening of purifying selection associated with the shift to selfing.
A major result of our analyses is that we need not invoke an extreme bottleneck of a single founder, rather the data are consistent with high levels of drift in a population with a small effective size potentially founded by a large number of individuals. 

The novel haplotype-based method developed herein allows us to partition polymorphism patterns between variation inherited from the ancestral outcrossing population 
	and new diversity introduced after the bottleneck.
By partitioning these sources of variation, our approach allows us to more clearly detail the relaxation in purifying selection associated with the transition to selfing. 
This partitioning also facilitates coalescent-based approaches to the demographic history of selfing populations and can therefore help infer the extent of a founding bottleneck, 
	identify population subdivision, and document recent population growth and geographic spread.
Therefore, beyond the application to \emph{Capsella}, the framework developed here can be used in other pairs of outcrossing/selfing species in order to build 
	a broad comparative view of the shift from outcrossing to self-fertilization. %MP rewording 
More generally, the ideas developed herein could be applied to many recently diverged species pairs in which one has gone through an extreme demographic bottleneck, leaving only a few recognizable founding haplotypes, regardless of mating system. 
%the diversity of processes underlying
\newpage
%%%%%%%%%%%%%%%%%%%%%%%%%%%%%%%%
% RESULTS
%%%%%%%%%%%%%%%%%%%%%%%%%%%%%%%%
\section*{Results}
%%%%%%%%%%%% SAMPLES / SEQUENCING %%%%%%%%%%
\subsection*{Samples / Sequencing}
% SEQUENCES %	
	{\bf Sequence data:}  We analyze SNP data generated from the transcriptomes of 11 \emph{Capsella} samples (six \rube{,} five \grand{}) aligned to the \rube{} reference genome \cite{Wright2012}. 
	SNPs were called using the GATK pipeline and subjected to an additional  series of quality controls (described in the \emph{METHODS}). 
	These calls were validated by comparison to 53 Kb of Sanger sequencing that overlapped a subset of these data, revealing highly replicable genotype calls across technologies and nearly identical values of $\pi$ (see \emph{METHODS}, and Table S1).  
	Throughout the paper we focus on detailing variation at four- and  zero- fold degenerate sites 
		(i.e. synonymous, and nonsynonymous sites), which we signify with the subscripts, $S$ and $N$ respectively.
	
	Together, our data span 124.6 Mb of the \rube{} genome,  covering 25,000 unigenes. 
	Of this 124.6Mb approximately 96\% could be assigned a recombination
		rate from a genetic map (map length = 339 cM) that was constructed
		from a QTL cross between \rube{} and \grand{} \cite{Slotte2012}. 
	While this genetic map may not be representative of that in \rube{,}
		it is more appropriate to measure haplotype lengths on a
		genetic rather than physical map, because the former provides information about the number of outcrossing events since coalescence, and so we quote both measures.

%	\yb{STEPHEN - DO YOU KNOW HOW MANY UNIGENES? I COULDN'T TELL IF IT WAS 25K or 4K FROM THE GENOME PAPER}

% SAMPLES %	
	{\bf Samples:}  	
	Our six \emph{C. rubella} samples consist of three plants from Greece, the native range of \grand{,}  and the putative location of the origin of \rube{} \cite{Foxe2009, Guo2009}, 
		and three from outside of Greece (Italy, Algeria, and Argentina), outside of \emph{C. grandiflora}'s range.
	We often partition our analysis into these two \emph{C. rubella} groups because
	Greek samples are likely closer to demographic equilibrium and have the opportunity to introgress with \grand{,} 
		while Out-of-Greece samples provide us with an opportunity to explore the influence of \emph{C. rubella}'s geographic expansion on patterns of sequence diversity.
%\yb{	PERHAPS A QUCIK SENTENCE JUSTIFYING SMALL N FOR THE AFS}

%%%%%%%%%%%% WHOLE GENOME SUMMARIES %%%%%%%%%%
\subsection*{Whole genome summaries}
Before presenting our haplotype-based analyses, we briefly summarize
patterns of sequence variation within and among species. 
These results, which are consistent with previous analyses 
	and are strongly concordant with Slotte \emph{et al.}'s \cite{Wright2012} analysis of the same data,  
	are summarized here for completeness. 
To generate empirical confidence intervals, 
	we calculate the upper and lower 2.5\% of tails of focal 
	summary statistics by resampling $250$ kb blocks with replacement.

{\bf Patterns of diversity and divergence:} 
In Table 1 and Figure S1A we show variation within and between populations and species. 
Interspecific divergence at synonymous sites ($d_S$) slightly
        exceeds synonymous diversity ($\pi_S$) in {\em  C. grandiflora}.
In turn, both of these estimates dwarf diversity in {\em C. rubella}. 
 Sequence diversity in \emph{C. rubella} is geographically structured, with pairs of Out-of-Greece samples being 	much more similar to one another than are Greek sample pairs (estimated as $\pi_S$), while pairs consisting of 	
	one Greek and Out-of-Greece sequence differ slightly more.
%Sequence diversity in \emph{C. rubella} shows the geographic structure of genetic diversity -- $\pi_S$ 
%	between pairs of Out-of-Greece samples is much lower than $\pi_S$ among Greek samples,
%	which is slightly less than $\pi_S$ between Greek and Out-of-Greek samples. 
The spatial structure of genetic variation in
        \emph{C. rubella} argues against recent introgression 
	between\emph{C. grandiflora} and sympatric Greek \emph{C. rubella}, 
	since divergence ($d_S$) between them is not significantly different from that between allopatric Out-of-Greece
          \emph{C. rubella} and \emph{C. grandiflora}. To further test this, we calculated
          a formal test of introgression, the $f_3$ statistic \cite{Reich2009,Patterson2012}, 
          which provided no evidence for introgression (see Text S1).

Additional characteristics of these data, specifically an excess
  of intermediate frequency variants and a relative excess of
  nonsynonymous variation, likely reflect genomic consequences of the transition to selfing. 
 For example, we observe an excess of intermediate frequency variation in both Greek
	and Out-of-Greece  \emph{C. rubella} samples as compared to constant
	neutral population expectations, consistent with a historical
	population contraction (Figure S1C-D).
A relaxed efficacy of purifying selection in \emph{C. rubella} is suggested by the level of nonsynonymous relative to synonymous variation within and between species (Figure S1B) --  
$\pi_N/\pi_S$ within \emph{C. rubella} is large (0.173) compared to both
	$\pi_N/\pi_S$ within \emph{C. grandiflora} (0.144), 
	and to $d_N/d_S$ or between species (0.146).

	{\bf The genomes of {\em C. rubella} individuals are largely autozygous:}
	Since \emph{C. rubella} is predominantly self-fertilizing, we expect most of an individual's genome to be autozygous -- 
		that is, an individuals two  chromosomes are predominantly identical by descent due to a very recent common ancestor.
	As expected, most \emph{C. rubella} individuals are homozygous at the majority of sites 
		(\emph{C. rubella} individuals are homozygous at 89\% to 95\% of non-singleton synonymous polymorphisms in \emph{C. rubella}, as compared to  \emph{C. grandiflora} individuals 
		who are homozygous at 55\% to 64\% of non-singleton synonymous polymorphisms), 
		likely due to numerous consecutive generations of self-fertilization in \emph{C. rubella}.
	However, some individuals contain a few genomic regions that are putatively allozygous, as manifested by high local levels of heterozygosity.
	Such regions have yet to be homogenized by selfing since the most recent ancestral outcrossing event,
		and are clearly demarcated and easily identified by higher levels of individual heterozygosity than in the rest of the genome (see Text S1 and Figures S9A-F). 
	In total, we infer that on average 7\% of a \emph{C. rubella} individual's genome is allozygous. 
To simplify our haplotype-based analyses, we ignore these
allozygous regions, which allows us to  directly observe the phase of
nucleotide variants. 	
In the \emph{METHODS} and Figure S8 we show that these excluded
allozygous regions do not contain unusual patterns of sequence 
diversity, and so their exclusion is unlikely to affect our inference (see \emph{METHODS}).

	%implies 
%		because in autozygous regions we can directly observe the  phase of
   %             nucleotide variants for most of the genome. 
%	Therefore, for the rest of our analyses, which are haplotype-based, we treat the small number of allozygous regions of an individual's genome
 %       as missing data to avoid the complications of having to
 %       resolve phase. Doing so does not appear to bias us away from unusually
 %       diverse regions, (see \emph{METHODS} and Figure S8). 
%	Additionally, the autozygous nature of the genome allows for an additional step of quality control, since the rare, isolated heterozygous sites in putatively 
%		autozygous regions are likely errors, and so we mask them and treat them as missing data (\emph{METHODS}).

%moved the bit about removing from hap analysis down to hap section

	%%%%%%%%%%%%%%%%%%%%%%%%%%%%%%%%%%%
	%%%%%%%%%%%%%%%%%%%%%%%%%%%%%%%%%%%
	%%%%%%%%%%%%%%%%%%%%%%%%%%%%%%%%%%%
	%%%%%%%%%%%%%        HAP       %%%%%%%%%%%%%%%
	%%%%%%%%%%%%%%%%%%%%%%%%%%%%%%%%%%%
	%%%%%%%%%%%%%%%%%%%%%%%%%%%%%%%%%%%
	%%%%%%%%%%%%%%%%%%%%%%%%%%%%%%%%%%%

	\subsection*{Comparisons within and among founding haplotypes}
	%%%%%%%%%%%%%%%%%%%%%%%%%%%%%%%%%%%
	%%%%%%%%    BASIC      HAP     SUMMARY    %%%%%%%%%%
	%%%%%%%%%%%%%%%%%%%%%%%%%%%%%%%%%%%

We now describe our novel haplotype-based analysis, which focuses on
	identifying haplotypes that founded  \emph{C. rubella}. 
By identifying these distinct founding haplotypes,
	we can divide variants in the extant \emph{C. rubella} population
	into those present in its founding lineages and new mutations. %variants that arose later. 
This information will allow us to infer a coalescent based model of the recent
	demography of \emph{C. rubella}.

	  {\bf {Identifying {\em C. rubella}'s founding haplotypes:}}
Figure 1 illustrates our approach to identifying \emph{C. rubella}'s distinct founding haplotypes, 
	a framework which will likely apply to many recently evolved selfing species. %across the genome in a recently derived selfing population, like \rube{}.  
At a given locus, all extant individuals trace their ancestry to one
	of a small number of founding lineages (which, for brevity, we call
	`founding haplotypes') that survive to the present (Figure 1A). 
These founding haplotypes should persist for long genetic map distances, 
	given the recent origin of \emph{C. rubella} and low effective recombination rate under selfing \cite{Nordborg2000}.

We define founding haplotypes as distinct \emph{C. rubella} lineages that do
	not share a common ancestor until they are present in the 
	population ancestral to \emph{C. rubella} and \emph{C. grandiflora}. 
A common way that this could occur is from the incomplete sorting of ancestral variation (Figure 1A). 
While a founding haplotype could, in principle, be
	introduced via introgression from \emph{C. grandiflora}, the lack of evidence 
	for introgression (above) suggests that this is rare.
While we observe no evidence for recent introgression, we note that our inferences, 
	with the exception of the coalescent modeling later, do not rely on assuming that introgression is rare.

%We caution that these founding haplotypes may recombined po occassionally 
%	or may represent multiple recombination events in a small genomic region.  
%However,  in some genomic regions  haplotype may have been introduced to \emph{C. rubella}   via introgression from \emph{C. grandiflora} 
	%(although, as we report above, we  see no strong evidence of introgression). 
%Furthermore, not all inferred founding haplotypes  need represent contiguous haplotypes actually present in the founders; recombination early in \emph{C. rubella}'s history could have created a chimeric haplotype.

\begin{figure}   
                \begin{center}   
               \includegraphics[width = 6in]{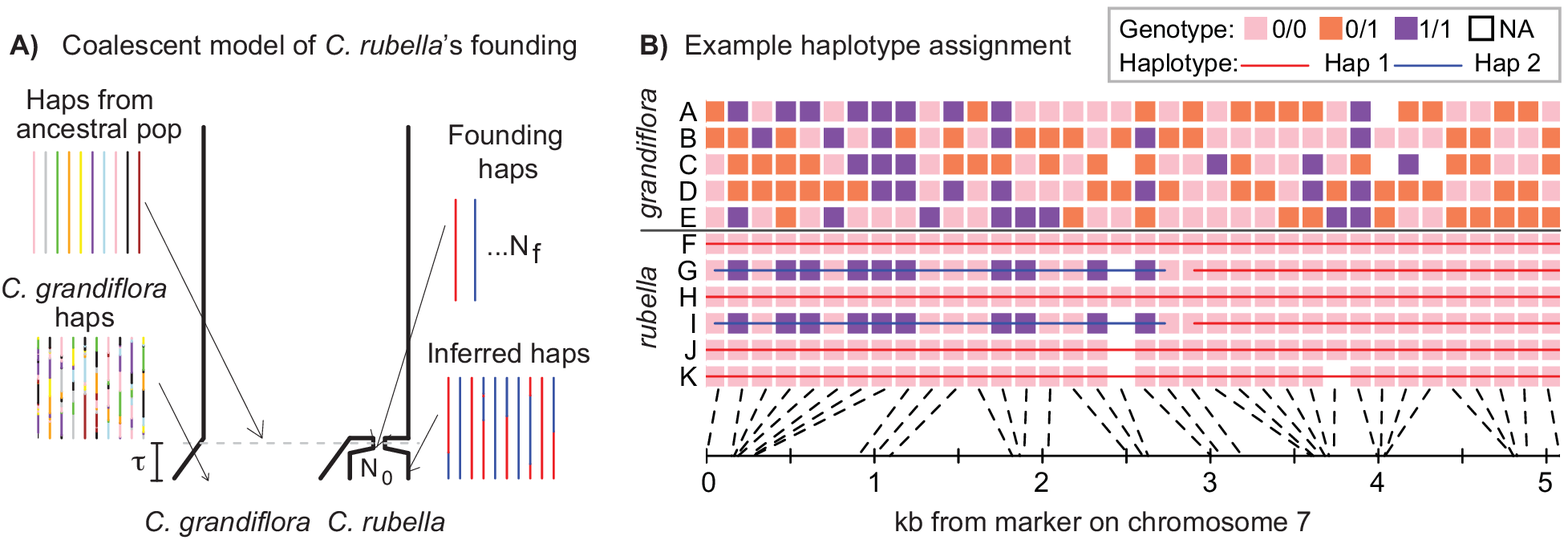}
                \end{center}
                \caption{{\bf The founding of \emph{C. rubella} and the identification of its founding haplotypes:}
                \emph{A)} A cartoon coalescent model of \emph{C. rubella}'s origin. 
                At time, $\tau$, a population ancestral to
                \emph{C. rubella} is formed by sampling $N_f$
                chromosomes (i.e. haplotypes, haps) from a large outcrossing population ancestral to both species, 
                and this selfing population quickly recovers to size, $N_0$.
                Because some of the $N_f $lineages are lost to drift, we can identify the founding haplotypes surviving to the present, which we color in red and blue.
                While recombination  scrambles ancestral chromosomes in \emph{C. grandiflora}, 
                	the low effective recombination rate in \emph{C. rubella} ensures that large chunks of founding haplotypes remain intact.
                \emph{B)} We aim to identify these founding haplotypes by using patterns of sequence variation (see text and \emph{METHODS} for details of our algorithm).
               Here, we present an example of founding haplotype identification in a typical genomic region. 
               To aid visualization, we label the major allele in \emph{C. rubella} as `0', and the allele that is rare or absent \emph{C. rubella} as `1', and only display genotypes at sites with common variants in  \emph{C. grandiflora}.
                In the left hand side of Figure 1B, there are clearly two distinct founding haplotypes on the basis of patterns of variation at sites polymorphic in both species. 
                On the right hand side, all \emph{C. rubella} individuals are identical at sites polymorphic in \emph{C. grandiflora}, so we infer a single founding haplotype.  
                }
                \label{Coal}
            \end{figure}

Using the model in \figref{Coal}A, we develop a non-parametric framework to robustly identify the genomic regions where  two \emph{C. rubella} individuals both have the same founding haplotype, versus two different ones (see \figref{Coal}B for an example, and \emph{METHODS} for more details).
A pair of individuals must have different founding haplotypes in genomic regions where they differ at multiple sites that are polymorphic in both species (assuming no recurrent mutation). 	
%We reason that a pair of individuals must have inherited distinct founding haplotypes %	where they differ at sites polymorphic in both %	species (assuming no recurrent mutation). 
We therefore assign pairs of individuals to distinct founding haplotypes in genomic regions 
	where they consistently differ at sites segregating in both species, and to the same founding haplotype 
	where they are identical at such sites (see for example the left portion of  \figref{Coal}B). 
Also,  in stretches of the genome where a number of sites are polymorphic in \emph{C. grandiflora} but fixed in \emph{C. rubella}, we assign all \emph{C. rubella} individuals to the same founding haplotype (see for example the right portion of Figure 1B, see \emph{METHODS}). 

	%Missing genotypic data for one or a subset of individuals can mislead a naive haplotype assignment. % algorithm, and potential result in mistaken founding haplotype labels. 

%	Thus, after initial pairwise assignment of sequences to the same or different founding haplotypes, 
To ensure robust founding haplotype calls we identified `ambiguous'
	genomic regions, where the assignments for different pairs of \emph{C. rubella} individuals reveal conflicts 
	(e.g. for three individuals, A, B, and C,  A=B, B=C, A$\neq$ C, due to missing data, where = and $\neq$ refer to the same or different preliminary haplotype assignment, respectively). 
Because haplotype calling in such `ambiguous' regions is
        problematic, we exclude them from our analysis (and return to
        discuss these genomic regions later).
In the \emph{METHODS} we describe these algorithms fully, with details of the number of SNPs and physical 
	distances that we require to assign samples to the same or different founding haplotypes. 
In  Text S1 we show that our results are robust to these cutoffs. 
	
	%Patterns of pairwise founding haplotype sharing provide a view of population history and a measure of relatedness between individuals. 

{\bf Patterns of pairwise founding haplotype sharing.} 
Figure 2A shows the proportion of the genetic map for which two
	\emph{C. rubella} individuals are assigned to the same founding
	haplotype (on average 72\%), distinct founding haplotypes (15\%), or for which haplotype assignment is ambiguous (13\%).
Figure S2 shows similar results measured by proportion of the physical map, and Figure S3 shows the robustness of these results to haplotype assignment cutoffs.   
In total, pairs of individuals transition between the same to different founding haplotypes between $500$ and $1000$ times, depending on the comparison.
Therefore, the haplotype-based analyses, below reflect at least $500$, and likely many more, different coalescent events per pair of individuals.

\begin{figure}	
			\begin{center}	
			\includegraphics{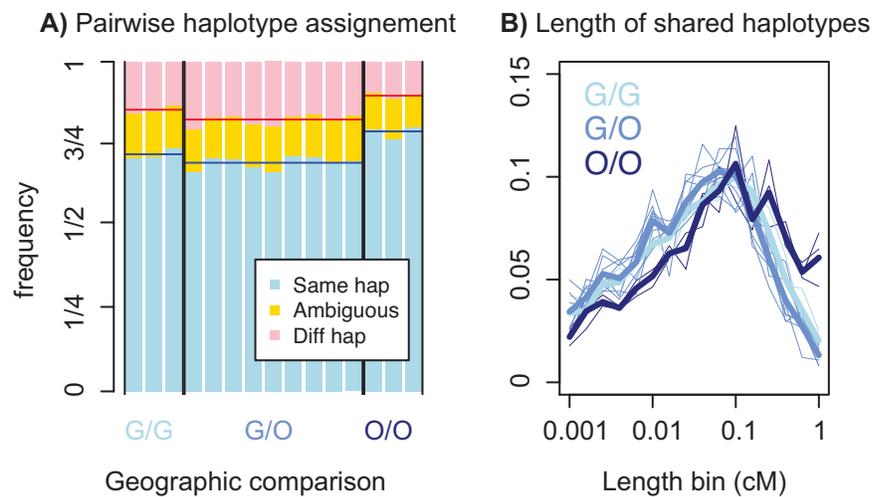}
			\end{center}
			\caption{{\bf Patterns of founding haplotype sharing in \rube{:}}  %$A$) The haplotype frequency spectrum in noncontroversial genomic regions homozygous in all individuals. 	
			\emph{A)} The proportion of the genome for which two individuals have inherited
				 the same or different founding haplotypes, or for which haplotype calls are ambiguous (see text for explanation).	
			The geographic origin of the pair is  denoted by G (Greek), or O (Out-of-Greece), e.g. a comparison between a Greek and Out-of-Greece pair is denoted by `G/O'.
			\emph{B)} The length distribution of regions assigned to the same founding haplotype. 
			Thin lines represent pairwise comparisons and thick lines represent mean values for this pairwise measure within a geographic class.
			In \emph{Text S1}, we recreate this
                        figure  utilizing physical, rather than
                        genetic distances, and find qualitatively
                        similar patterns (Figure S2).}
			\label{HFS}
		\end{figure}

As expected, assignment of pairs of samples to the same or different founding haplotype is consistent with patterns of pairwise sequence diversity reported above.
Out-of-Greece pairs are assigned to the same founding haplotype more often than pairs from Greece,
	 and comparisons between a Greek and Out-of-Greece plant have the lowest proportion of founding haplotype sharing.
The same  pattern is reflected in the length distribution of founding haplotype blocks (\figref{HFS}B). 
	%Additionally, regions of founding haplotype sharing of  Out-of-Greece pairs are longer than pairs from Greece, while Greek/Out-of-Greek pairs have slightly shorter runs of founding haplotype sharing than either within-region type of pair. (\figref{HFS}B).  
This high level of founding haplotype sharing suggests that there has
	been extreme drift during or subsequent to the founding of
	\emph{C. rubella}, particularly outside Greece.

		\subsubsection*{Patterns of polymorphism within and between founding haplotypes}

	%%%%%%%%%%%%%%%%%%%%%%%%%%%%%%%%%%%
	%%%    VARIATION    WITHIN   AND   BETWEEN   HAPS     %%%%
	%%%%%%%%%%%%%%%%%%%%%%%%%%%%%%%%%%%

We next used these founding haplotype designations to partition
	patterns of polymorphism. 
We denote comparisons between individuals
	assigned to the same founding haplotype in a genomic region, averaged across
	all such regions genome-wide, by the phrase, `within founding haplotypes'. 
In turn, we denote comparisons between individuals assigned to different founding haplotypes,
	averaged across all such regions genome-wide, by the phrase, `among founding haplotypes'.
As above, the subscripts N and S refer to synonymous and on synonymous sites, respectively. 
To provide empirical 95\% confidence intervals for reported statistics, 
	we resample regions of haplotype assignment with replacement.
	% , and report the extreme $2.5\%$ tails in square brackets and/or as error bars in figures.

%Again we resample with replacement to generate an empirical $95\%$ confidence interval, which we state in brackets. 
%In these haplotype-based analyses we resample at the level of founding haplotype.
	%%%%%%%%%%%%    PAIRWISE   PI   %%%%%%%%%%%%%

{\bf Diversity within founding haplotypes is low, diversity among founding haplotypes is high:}	

\begin{figure}	
			\begin{center}	
			\includegraphics[width = 6in]{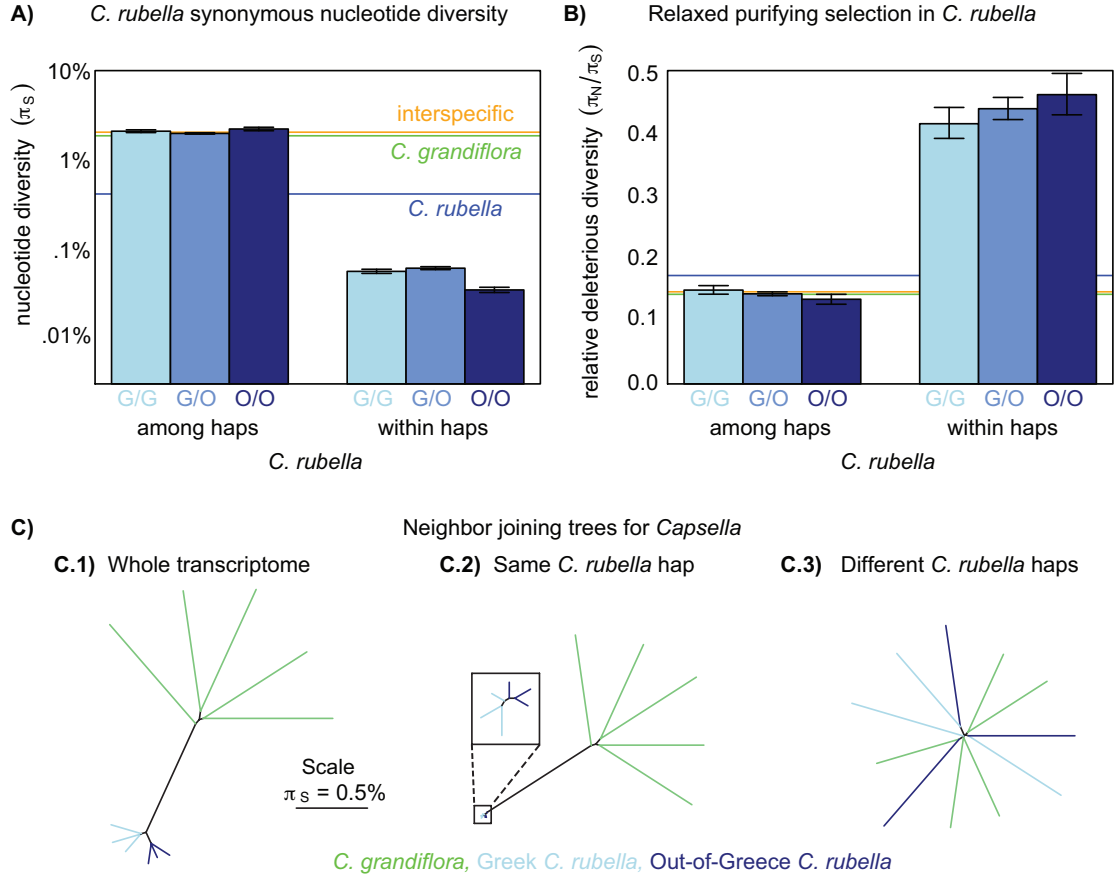} 	
			\end{center}
			\caption{{\bf Variation within and among \emph{C. rubella}'s founding haplotypes:} 
				\emph{A)} Pairwise nucleotide
                                diversity ($\pi_S$) within
                                and among \emph{C. rubella}'s founding
                                haplotypes at synonymous sites (see
                                Table S3 for values). 
				\emph{B)} Ratio of  nucleotide
                                diversity at non-synonymous relative
                                to synonymous sites ($\pi_N/\pi_S$) within and among \emph{C. rubella}'s founding haplotypes. 
				Error bars mark the upper and lower 2.5\% tails
				and are generated by resampling blocks assigned to different (left hand side) or same (right hand side) founding haplotypes.  
				In the top panel (A and B), orange, green, and blue horizontal lines are drawn for reference to interspecific comparisons, 
					comparisons within
                                        \emph{C. grandiflora}, and
                                        genome-wide \emph{C. rubella}
                                        comparisons, respectively
                                        (taken from Table 1). 
				\emph{C)} Neighbor joining trees in
                                \emph{Capsella}, using all comparisons
                                (\emph{C.1}), comparisons within
                                (\emph{C.2}), or among (\emph{C.3})
                                founding haplotypes to generate entries in the pairwise distance matrix for comparisons within \emph{C. rubella}.
				All distances are generated from nucleotide diversity at synonymous sites.
				}
		\label{pishap}
		\end{figure}

For pairs of \emph{C. rubella} samples, we estimated $\pi_S$ in genomic regions assigned to the same or different founding haplotypes. 
Regardless of the geographic origin of the \emph{C. rubella} plants analyzed, $\pi_S$ among haplotypes is similar to estimates of interspecific diversity (Figure 3A). This suggests that our inferred founding haplotypes correspond well to \emph{C. rubella}'s founding lineages. 
%Diversity among founding haplotypes does not differ from interspecific divergence and does not depend on the geographic comparison (Figure \ref{pishap}A)  --  %($d_S = 2.03\%$
%	results that are consistent with the idea that the founding haplotypes we infer correspond well to \emph{C. rubella}'s founding lineages. 
By contrast, diversity within founding haplotypes is very low -- approximately an order of magnitude lower than baseline diversity in this inbred species (Figure \ref{pishap}A). 
Additionally, the amount of variation within founding haplotypes depends on the geographic location of samples.
As in genome-wide summaries, diversity	within founding haplotypes is highest across geographic comparisons,
		lowest in Out-of-Greece pairs, and intermediate within Greece pairs (Figure \ref{pishap}A).
All of these results are robust to cutoffs for founding haplotype assignment (Figure S5A).  
Since variation within founding haplotypes must have arisen since
	\emph{C. rubella}'s founding, this paucity of variation
	could reflect either little time to accrue novel mutations, 
	or a small effective population size limiting \emph{the extent of} variation.  
%Below, we provide support of the latter hypothesis.
Below, we show that the small effective population size 
  explanation is a strong explanation of these data.

%($ \pi_{W,S} = 5.29 [5.11,5.40] \times 10^{-4}$), 
%		even compared to genome-wide values for \emph{C. rubella} ($\pi_U = 40.72  \times 10^{-4 }$  \figref{pishap}{A}).

% ($\pi_{W,U}= 6.05 [5.78,6.24] \times10^{-4}$)   ($\pi_{W,S}= 3.48
% [3.20,3.66] \times10^{-4}$ $\pi_{W,S}= 5.55  [5.24,5.84]\times10^{-4}$
%\gc{GOT HERE, want to add sentence of expalantion of biology} 

These results offer a straightforward interpretation of \rube{} diversity across the genome as a  mosaic of relatively few founding haplotypes that have survived to the present day.
Thus, we expect sequence diversity to vary as we transition between genomic regions with different numbers and frequencies of surviving founding haplotypes.
Patterns of polymorphism are consistent with this view --  there is a strongly negative relationship between the frequency 
	of the most common founding haplotype and sequence diversity (Pearson correlation, $r^2 = -.233\text{ }[95 \% \text{CI } = -0.224, -0.243] $, see \figref{chr7}{ } and Figure S7).

To further aid visualization of the structure of variation within and among founding haplotypes 
	 we present a set of neighbor joining trees constructed from pairwise distance matrices  (\figref{pishap}C).
	 %, we present a set of neighbor joining trees to visualize the pairwise matrices representing variation within and among species and founding haplotypes. 
The tree constructed from the entire transcriptome (\figref{pishap}C.1)  shows 
	little genetic diversity within \emph{C. rubella}, 
	the distinctness of \emph{C. rubella} from \grand{,}
	and the clustering of Out-of-Greece \emph{C. rubella} samples. 
In contrast, \figref{pishap}C.2 reveals diversity within founding haplotypes is completely dwarfed by diversity within \grand{} and interspecific divergence; however, by zooming in on the \emph{C. rubella} branch of this tree we recover the clustering of Out-of-Greece samples  (top left of \figref{pishap}C.2). 
Comparisons among founding haplotypes reveal a starlike structure for all sequences (\figref{pishap}C.3).
 Because \emph{C. rubella} samples that have different founding haplotypes do not cluster with one another, this  
	suggests that \emph{C. rubella}'s founders were close to a random selection of ancestral variation,  
	rather than a distinct \emph{C. rubella} sub-population, and that there has been little allele frequency divergence genome-wide in \grand{} since the founding of \rube{.}

{\bf  Putatively deleterious variation is overrepresented within founding haplotypes:}  
The ratio of non-synonymous  to synonymous variation among
        \emph{C. rubella}'s founding haplotypes is low, %$\pi_{A,N}/\pi_{A,S} = 0.139 [0.137,0.141]$, 
         resembling that found in \emph{C. grandiflora}.
By contrast, much of the diversity within founding haplotypes is nonsynonymous (nearly one-third)
        %$($\pi_{W,N}/\pi_{W,S} = 0.438 [0.425,0.449]$) 
         (\figref{pishap}B), a result that is robust to founding haplotype calling cutoffs (Figure S5B). 
Since the excess nonsynonymous variation in \emph{C. rubella} is segregating within haplotypes, 
	and therefore novel, elevated nonsynonymous diversity in this species suggests a relaxation 
	in the efficacy of purifying selection following the transition to selfing  \cite{Glemin2007}.
 This elevated $\pi_N/\pi_S$ within compared to among founding haplotypes,
  is also reflected in patterns of variation at polymorphic sites
  private to a species sample. 
  That is, the ratio of nonsynonymous to synonymous polymorphisms unique to
our \emph{C. rubella} sample is 3.5 fold higher than this ratio in polymorphisms unique to
our \emph{C. grandiflora} sample.
Overall this shows that polymorphisms that have arisen since the
founding event within \emph{C. rubella} are strongly enriched for
non-synonymous, likely deleterious, variants.

\subsubsection*{The frequency of different founding haplotypes in \emph{C. rubella}:} % one of up to six
Building on pairwise founding haplotype assignments, we 
        identified distinct founding haplotypes across the \emph{C. rubella} genome. 
This higher-order haplotype assignment provides information about both 
	the frequency spectrum of founding haplotypes, 
	and the allele frequency spectrum within founding haplotypes.
		
To construct the set of founding haplotypes in a genomic region,
        we simultaneously evaluate all patterns of pairwise
        founding haplotype assignment in this region  (see \emph{METHODS} for a complete description of the algorithm). 
For example, in the left hand side of \figref{Coal}B, all
        pairwise comparisons between individuals F, H, J, and K show
        them to be identical at sites polymorphic in both species, and so they are
        assigned to haplotype 1 (indicated by red lines). 
	Similarly, individuals G and I are assigned to the same
        founding haplotype (haplotype 2, blue lines), which is
        distinct in pairwise comparisons from founding haplotype 1. 
	On the right hand side of this figure all \emph{C. rubella}
        samples are identical for a stretch of sites polymorphic in
        \emph{C. grandiflora}, and so are assigned to the same founding haplotype.

{\bf A summary of founding haplotype assignment:}
 Using these assignments, we find that for 57\% of the genome,
	\emph{all} \emph{C. rubella} individuals in our sample have inherited the same founding haplotype, 
	for 19\% of the genome, all individuals can be unambiguously assigned to one of two haplotypes, 
	and for 25\% of the genome at least one individual could not be unambiguously assigned to a founding haplotype. 
The fact that so much of the {\em C. rubella} genome contains so little diversity in founding haplotypes 
	suggests that either very few individuals founded  {\em C. rubella}, % with little subsequent gene flow from \emph{C. grandiflora}, 
	or that nearly all of the diversity present in a large  founding population has been lost by subsequent drift and selection.  
Below we use the frequency spectrum within founding haplotypes and coalescent modeling to distinguish between these possibilities.

{\bf Regions with more than two founding haplotypes.}
Overall $\sim$75\% the genomes of our \emph{C. rubella} samples can be
	unambiguously assigned to $\leq 2$ founding haplotypes. 
The remaining quarter of the genome is split between genomic regions with more than two founding haplotypes, 
	ambiguous haplotype assignment and/or transitions between haplotypes for at least one sample 
	(see Table S2 for the sensitivity of these results to haplotype calling cutoffs). 
%	Overall, slightly less than 10\% of the genome has evidence of more than two founding haplotypes in our sample, a point that we return to below.
Convincing evidence for even a single genomic region containing more
	than two founding haplotypes would rule out the hypothesis that the
	ancestry of \emph{C. rubella} can be traced to a single founder with
	no subsequent introgression \cite{Guo2009}. 
However, there are numerous alternative reasons why a small portion of the \emph{C. rubella} 
	genome may appear to contain more than two founding haplotypes. 
These explanations  include the misalignment of paralogous regions as well as incorrect founding haplotype assignments
	caused by multiple historical recombination events. % between founding haplotypes. 
We therefore carefully investigate the possibility that some genomic regions contain more than two founding haplotypes.

We identified genomic regions likely containing more than two founding haplotypes by a sliding window analysis moving across the
	genome of all trios of our six \emph{C. rubella} samples. 
In windows of $20$ sites with more than one copy of the minor allele in \emph{C. grandiflora}, moving
	one such SNP at a time, we noted candidate regions where each member
	of the trio differs from the others at one or more of these SNPs. 
We pruned this list of candidates in two ways.   
We  included only windows where each member of the trio is
	differentiated by $\pi_S > 1\%$ in the candidate region, a level much higher than that within founding hapotypes and within the range of diversity in \emph{C. grandiflora}, to  ensure
	that the windows likely include $3$ distinct founding haplotypes.   
To minimize the chance that such high diversity regions represent misassembly, we required that at least  one member of the trio is similar to another sample ($\pi_S < 0.5\%$) in that genomic region. 

We identified 172 genomic regions likely to harbor more than two founding haplotypes, and we present nine exemplary regions in Figure S10.
In total, such regions make up approximately 2\% of the genome. 
These regions are generally quite short (53 are 10 kb or less, 132 are
less than 20 kb, and all are shorter than 70 kb).
The length distribution of genomic regions with $> 2$ haplotypes likely reflects recombination since the origin of
	\emph{C. rubella}, and suggests that these additional founding haplotypes have probably not been recently introduced by introgression. 
Given their small size and our stringent criteria, we likely have underestimated the fraction of the genome
	with $>2$ founding haplotypes.

%While we believe this estimate is reasonable we believe it is imperfect as we likely missed some very short regions and may not have removed all potential sources of error.

\begin{figure}	
				\begin{center}	
				\includegraphics{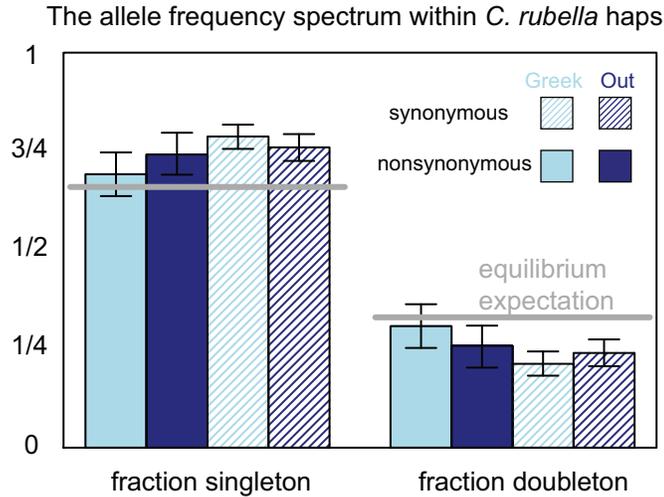} 	
				\end{center}
				\caption{{\bf The allele frequency spectrum within \rube{}'s founding haplotypes:} 
				The proportion of polymorphic derived alleles within a founding haplotype observed as singletons or doubletons, split by geography and synonymy. 
				Light and dark blue represent comparisons within Greek and Out-of-Greece samples, respectively.
				Filled and hatched bars represent synonymous and non-synonymous sites, respectively.  
				Error bars represent the upper and lower 2.5\% tails of the allele frequency spectrum when founding haplotypes are resampled with replacement.
				Grey lines represent expectations of a model for neutral mutations at mutation-drift equilibrium.
				}
				\label{afs}
			\end{figure}

\begin{figure}	
				\begin{center}	
				\includegraphics[width = 6in]{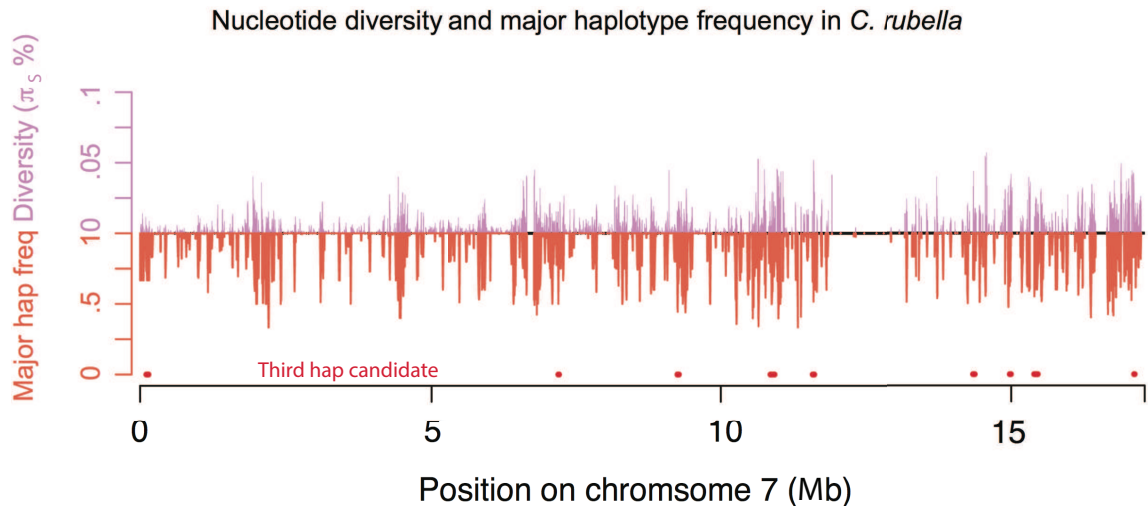}
			\end{center}
				\caption{{\bf Diversity across chromosome seven in \emph{C. rubella}:}
					Mean pairwise synonymous diversity (purple, upward pointing lines) and 
					major founding haplotype frequency (orange, downward pointing lines) across chromosome seven.
					 %$ \pi_U$ and the major haplotype frequency in such regions do not differ from randomly sampled, sized matched regions (not shown). 
					 Red points mark regions putatively containing more than two extant founding haplotypes. 
					 Values of $\pi_S$ and major
                                         founding haplotype frequency
                                         are averaged across
                                         overlapping sliding windows
                                         (ten kb windows with a two kb
                                         slide), here only windows
                                         with data for $> 1000$ sites of pairwise
                                         comparisons are
                                         evaluated. See Figure S7, for
                                         plots of all chromosomes.
					%{\em B and C)} Pairwise sequence diversity between three individuals for an exemplar portion of chromosome seven containing two ({\em B} ) or potentially three ({\em C} ) haplotypes.
					%Lines at $ \pi_S$ = 2\% are presented as a reference for expected diversity among haplotypes.
%					$C)$ Pairwise sequence diversity between three individuals all assigned to different haplotypes (beginning at position $10\,874\,000$) for a portion of chromosome seven.
				} 
				\label{chr7}
			\end{figure}

{\bf  No excess of high frequency derived alleles within {\em C. rubella} founding haplotypes:}
	We make use of the allele frequency spectrum within founding haplotypes to distinguish between two alternative models of \emph{C. rubella}'s origin -- an extreme but short-lived bottleneck at its origin or a long-term reduction in population size.
	Within founding haplotypes, the frequency spectrum in Greece resembles the expectation under a constant population size model, 
		and  there is only a slight excess of rare derived alleles outside of Greece (\figref{afs}), a result robust to the choice of cutoffs for the labeling of founding haplotypes (Figure S4). 
	Since diversity within founding haplotypes is close to its
        expectation under drift-mutation equilibrium, the low level of variation within founding haplotypes in Greece reflects a small long-term effective population size,  
        rather than solely the effect of a dramatic bottleneck at the founding of C rubella (we quantify this statement shortly through coalescent modeling). 
	The slight excess of singletons within haplotypes outside of Greece is consistent with an out-of-Greek expansion; 
		however, given the broad geographic sampling we cannot exclude the confounding effect of population structure \cite{Ptak2002}. 
%	These data argue against a major population contraction which has yet to reach migration-drift equilibrium, which would generate a strong excess of singletons. 

We also used the allele frequency spectrum to test alternative explanations of the excess of nonsynonymous variation within founding haplotypes. 
Specifically, this elevated $\pi_N/\pi_S$ could represent a relaxed efficacy of purifying selection in \emph{C. rubella}, 
		or may reflect a departure from demographic equilibrium 
		(whereby the excess of non-synonymous variants is due to the fact that 
		many of the variants in \emph{C. rubella} are young and hence at low frequency). 
However, the similarity of the allele frequency spectrum at  synonymous and non-synonymous sites within founding haplotypes (\figref{afs}) 
	 argues against a demographic explanation for elevated $\pi_N/\pi_S$
	 and suggests a weakening efficacy of purifying selection in \emph{C. rubella}, presumably cause by its reduced effective population size.

	%%%%%%%%%%%%%%%%%%%%%%%%%%%%%%%%%%%
	%%% DEMOGRAPHIC INFERENCE IN RUBE FROM HAPS  %%%%
	%%%%%%%%%%%%%%%%%%%%%%%%%%%%%%%%%%%

\subsubsection*{	Inferring the number of founders and the timing of speciation:}
So far, we have examined  patterns of diversity in \emph{C. rubella} with little reliance on specific models or assumptions. 
To complement these analyses, we build a coalescent-based framework to
	infer the parameters of a simple demographic model of \emph{C. rubella}'s history from the results above.
To facilitate this inference we introduce a few assumptions. 
The most restrictive of these is that introgression between {\em C. rubella} and {\em C. grandiflora} has been negligible. 
While we cannot rule out the possibility of infrequent and/or very old introgression events,  
	the similarity in divergence between {\em C. grandiflora} and both Greek (sympatric) and Out-of-Greece (allopatric) {\em C. rubella} populations,  
	and the positive $f_3$ statistic (\emph{Text S1})
	argue against recent common introgression. 
%Below we are careful to describe alternative interpretations of our findings in the face of substantial gene flow. 
Additionally, conversion of synonymous site diversity measures into a time-scale of years requires assumptions about the mutation rate, variation in this rate, and life-history.  
Following previous work on \emph{Capsella} \cite{Foxe2009,Guo2009}, we assume an average neutral mutation rate ($\mu$) of $1.5 X 10^{-8}$ 
	 per base per generation \cite{Koch2001} in both species and an annual life history, so that a neutral position in \emph{C. rubella}  experiences $\mu$ mutations per a year. 	
To change these rate assumptions, divergence times can be linearly rescaled by alternative estimates of $\mu$ and/or life history descriptions. 
For example, to use a more recent estimate of $\mu$, $7 \times10^{-9}$ \cite{Ossowski2010}, 
	we can simply multiply our estimates, below, by roughly a factor of two.

As a first estimate of the split date between \emph{C. rubella} and \emph{C. grandiflora}, 
	 we use levels of diversity within and between species to estimate a divergence time ($\tau$, following \cite{Hudson1987}). 
In addition to assuming no introgression, this model also assumes that the expected pairwise coalescent time in \grand{} is the same today as it was in the population ancestral to \grand{} and \rube{}. 
Under these assumptions, divergence at synonymous sites should
be given by $ d_S  \approx \pi_{S \text{\emph{ C. grandiflora} }}  +2 \tau \mu$. 
Solving for $\tau$ and substituting our estimates of $d_S$ and $\pi_S$ within \emph{C. grandiflora}, we estimate a split time of  $\tau = (0.0203 - 0.0186)/(2\mu) = 8.5 \times 10^{-4} / \mu \approx 56.5 \text{ky}$.

	{\bf Demographic model:} 	
The estimate above provides an approximate divergence date but no additional details about the founding of  \emph{C. rubella}.
We aim to build a model that captures the major demographic events in {\em C. rubella}'s history and makes use of the founding haplotype approach introduced in this manuscript.
Throughout, we limit this analysis to four exchangeable samples (three from Greece and one from Out-of-Greece), so that our inference is not misled by population structure \cite{Ptak2002}.
Unlike the divergence estimate above, this model is robust to both 
	introgression from \emph{C. rubella} into \emph{C. grandiflora}, 
     and to changes in \emph{C. grandiflora}'s effective population size,
	but assumes no introgression from \emph{C. grandiflora}  into \emph{C. rubella} in the last $\tau$ generations.

	Inspired by previous methods that aim to infer the number of founding chromosomes from patterns of genetic variation \cite{Anderson2007,Leblois2007}, 
		we use coalescent modeling to jointly estimate the number of founding chromosomes
                and the time of \emph{C. rubella}'s founding.
We use the model  (depicted in \figref{Coal}{A})  where \emph{C. rubella} was founded
  	$\tau$ generations ago by a founding population of $N_f$ ($N_f = 2 N_{\text{founding individuals}}$)
	founding chromosomes, which instantly grew to its current effective population
	size of $N_0$ chromosomes. 	
We infer the parameter, $N_f$, and the compound parameter of the
	population-scaled founding time $\tau/N_0$ in a composite likelihood
	framework (see \emph{METHODS} for full details).  
To do so, we generate expected values of the allele frequency spectrum within founding haplotypes and the fraction of genomic windows
	where all samples inherited the same founding haplotype by simulating a coalescent model across a grid of $N_f$ and $\tau/N_0$. 
We then compute the composite likelihood of these aspects of our data across a grid of $\tau/N_0$ and $N_f$, 
	and resolve the compound parameter, $\tau/N_0$, by including information contained in diversity within founding haplotypes.  
In \emph{Text S1} we show that our inferences are
	robust to the choice of cutoffs for the labeling of founding haplotypes (Figure S6).

\begin{figure}	
				\begin{center}	
				\includegraphics[width=6in]{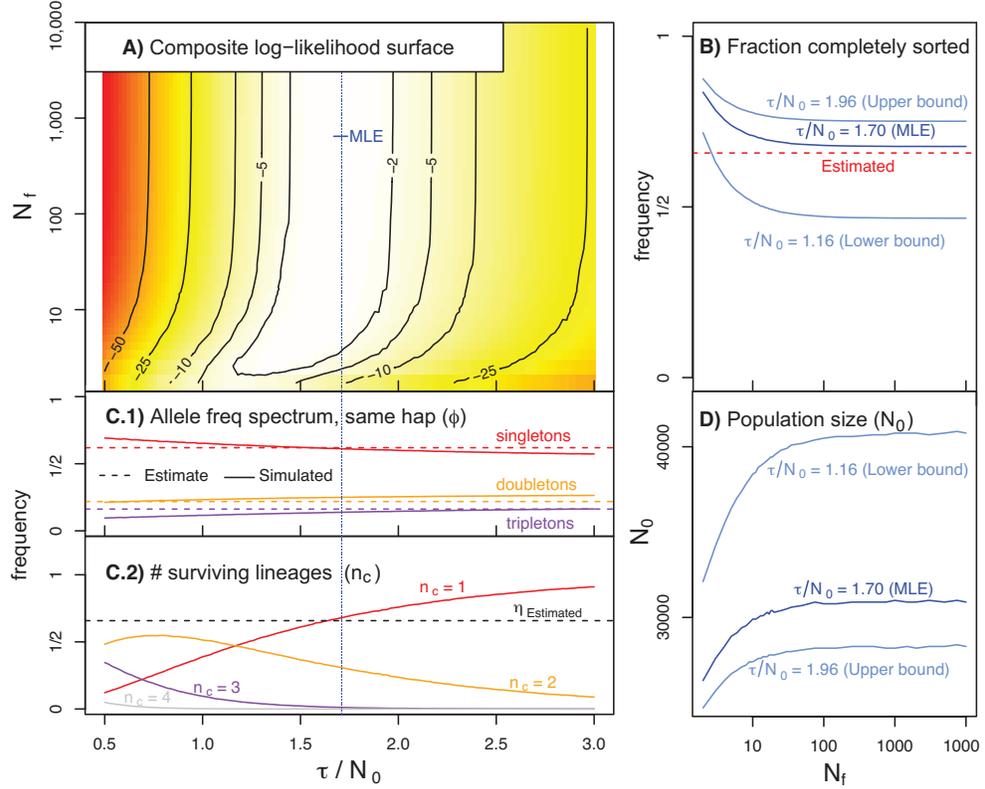} 	
				\end{center}
				\caption{{\bf A summary of our
                                    coalescent model of the history of \rube{:}} 
				\emph {A)} The relative composite
                                log-likelihood surface as function
                                of $\tau/N_0$ and $N_f$.
				\emph{B)} The probability that all individuals coalesce to the same founding haplotype ($\eta$) as a function of $N_f$ and three estimates of values $\tau/N_0$ 
					(the MLE, lower and upper confidence intervals). 
				The dotted red line indicates the value of ($\eta$) directly estimated from the data.
				\emph{C)} A summary of simulation results (assuming $N_f = 1000$).
				\emph{C1)} The frequency of singletons, doubletons, and tripletons observed in simulation (full lines), and estimated from 
					our data (dashed lines)
                                        conditional on all four
                                        samples deriving from the same
                                        founding haplotype.
				\emph{C2)} The frequency of one, two,
                                three or four lineages surviving to
                                the founding event. 
				When $N_f$ is large, $Pr(n_c) =1$ is the probability that all samples coalesce to the same founding haplotype. 
				The dotted black line portrays the
                                estimated frequency of all four
                                samples residing on one founding
                                haplotype
                                $\eta_\text{estimated} = 0.66$.
				\emph{D)} The estimated current effective number of chromosomes in \rube{} ($N_0$) as a function of the number of founding chromosomes ($N_f$).
				We plot this  for three different values of  $\tau/N_0$ (the MLE, as well as the lower and upper confidence intervals).
				 These results are robust to haplotype
                                 labeling criteria in
                                 \figref{instantgrowth}{}  (see
                                 \emph{Text S1, Figure S6}).
			}   
				\label{instantgrowth}
			\end{figure}

Our likelihood surface with respect to \emph{C. rubella}'s population-scaled founding
	time ($\tau/N_0$) shows a strong peak at a relatively large value of  $\tau/N_0$ (MLE
	= 1.7, with two log likelihoods  confidence interval of $ 1.2 < \tau/N_0 < 1.9$, \figref{instantgrowth}{A}). 
 This reflects the frequency with which all individuals inherit the same founding haplotype (\figref{instantgrowth}{B}), the  slight excess of singletons within 
	founding haplotypes, and the preservation of alternative founding haplotypes in \emph{C. rubella} (\figref{instantgrowth}{C}).  
Given this estimated range of $\tau/N_0$, we resolve this compound parameter by using our estimate of diversity within founding haplotypes (see \emph{METHODS}). 
Doing this, we infer the current effective number of 
	chromosomes, $N_0$ (\figref{instantgrowth}{D}), to lie between 
	25000 and 42000, and a split time, $\tau$, between 48 and 52 kya. 
This range is reasonably consistent with our estimated split 
	time of 56 kya obtained using a relatively independent source of  
	information %available in $\pi_S$ in \emph{C. grandiflora} and  $d_S$ 
	(see above).  
Our likelihood surface shows a long ridge in parameter space
        with respect to the number of founding chromosomes ($3 \leq N_f < \infty$ two log-likelihood confidence interval).
Therefore, while our data are consistent with few to many founding individuals,  a single founder is particularly unlikely.

	%%%%%%%%%%%%%%%%%%%%%%%%%%%%%%%%%%%
	%%%%%%%%%%%%%%%%%%%%%%%%%%%%%%%%%%%
	%%%%%%%%%%%%%%%%%%%%%%%%%%%%%%%%%%%
	%%%%%%%%%        DISCUSSION        %%%%%%%%%%%%%%
	%%%%%%%%%%%%%%%%%%%%%%%%%%%%%%%%%%%
	%%%%%%%%%%%%%%%%%%%%%%%%%%%%%%%%%%%
	%%%%%%%%%%%%%%%%%%%%%%%%%%%%%%%%%%%
				
\section*{Discussion}

We present a novel framework to interpret patterns of sequence diversity in  recently founded populations by viewing the genome as stretches of ancestry inherited from distinct founding chromosomes. 
We exploit this view to provide a detailed characterization of the evolutionary transition from outcrossing to selfing in \rube{.}
In principle, our conceptual approach is applicable to any founding event recent enough to preserve a reasonable portion of polymorphism present in the founders, regardless of mating system.
The application to  \emph{Capsella} was aided by the fact that few founding lineages contribute ancestry to our \rube{} sample, and that levels of linkage disequilibrium differ so starkly between \rube{} and \grand{,} making identification of the founding haplotypes relatively easy.
As these criteria are met by many recently founded selfing species and populations (e.g within \emph{Leavenworthia}, \emph{Mimulus spp.}, \emph{Arabidopsis lyrata}, and \emph{Clarkia  xantiana}   \cite{Sweigart2003,Wu2008 ,Mable2007,Busch2011, Pettengill2012}), including a number of commercially important species (e.g. indica rice and soybean \cite{Caicedo2007, Lam2010}), our framework should be of broad use as population genomic resources continue to be developed in these systems \cite{Branca2011,Ness2010,Ness2011}.

Our approach provides a new way of thinking about patterns of nucleotide diversity across the genomes of  recently derived selfers.
Moving across two phased genomes, we transition between regions in which our samples coalesce at or since the origin of selfing, and regions in which samples do not coalesce until they join the ancestral outcrossing population. 
Critically, we can use polymorphism present in a proxy for the outcrossing progenitor population (\emph{C. grandiflora}) to assess if two individuals have inherited the same or different founding haplotypes, since individuals that differ at ancestrally segregating sites almost certainly inherited different founding haplotypes.

\paragraph{Concerns about samples sizes. }

%One concern about our analyses is the small sample sizes. 
While we have sequence data from only six \emph{C. rubella} samples 
	(and often make use of three to four genomes to control for population structure), 
	these transcriptomic data  provide information about hundreds to thousands of
	genealogical histories as we move along the genome. 
Therefore the small number of sequenced individuals provides plentiful information
	about population history. 
A recent demonstration of this principle  is the development of coalescent 
	methods to infer population history from a single individual's genome  \cite{Li2011}. 

In particular, our findings about the small number of founding
haplotypes are likely generalizable to the population, since much of the common diversity (i.e. that contained in the deep parts
of the genealogy) in large samples is expected to be found in small samples \cite{Wakeley2006}.
This view is supported by the consistency of our findings and those of Guo et al. \cite{Guo2009},
	 who usually found one or two distinct haplotypes at each of 17 loci in a survey  25  {\em C. rubella} individuals.

While it is likely that our analyses, based on small sample sizes,
	have captured many aspects of the founding of \emph{C. rubella}, 
	larger samples will provide a fuller view of recent events.  
For example, additional genome-wide samples would provide 
	access to lower frequency variants (i.e. more novel mutations), 
     providing information about more recent
     population growth \cite{Adams2004, Keinan2012}, 
	and finer resolution of population structure. 
 Additionally, sequence data from more individuals would provide a finer resolution to the frequency spectrum of ancestral polymorphisms, and would help clearly identify genomic regions with more than two founding haplotypes. 
Therefore, additional samples could facilitate a more refined view of \emph{C. rubella}'s initial founding, and could potentially narrow the confidence intervals on our estimates of founding time, population growth rates, and population size.

\paragraph{A new view into the history of \emph{Capsella rubella}}

Our haplotype-based approach provides a  rough characterization of the  history of the selfing species, \rube{.} 
We note that since we have sequence data for only a handful of samples, we cannot provide fine resolution of recent demographic events in the history of the species.  
Assuming a mutation rate of $1.5 \times 10^{-8}$ \cite{Koch2001}, 
	we infer that approximately 50 kya,  %(potentially during the `long mild middle' of the last ice age), [REMOVED PARANTHETICAL]
	a \grand{}-like ancestral population of unknown size became largely selfing and gave rise to \rube{.}
Much of the ancestral diversity present in the founding population has since been lost due to subsequent drift and selection. % since \rube{}'s origin.  
In fact, two \rube{} individuals inherit different founding haplotypes for on average only $\approx 20\%$ of their genome. 
Despite this, the diversity maintained from the founding population
makes up roughly 90\% of extant pairwise sequence diversity in
\rube{,} since little diversity has arisen since its founding.  
We now turn to discuss some of the specifics of the founding and
  subsequent history of \emph{C. rubella}.
%\gc{do we need a little calc in methods on this?} 

%	However, as  small samples contain much of the common diversity of larger samples, i.e. much of the genealogical information about deeper events, 
%		larger samples are unlikely to provide a much finer resolution of the founding of  \emph{C. rubella}. 

%Since its origin, \rube{} has experienced a long-term reduction in population size as compared to is outcrossing progenitor, \grand{,} and therefore maintains less diversity and relatively more non-synonymous diversity than \grand{.} 
%And sometime between now and the origin of \emph{C. rubella}, a population left Greece and spread across the globe. 
%WE TALK ABOUT THE REST LATER

\paragraph{No obvious signal of an extreme bottleneck:}
%For example, because selfing plants are largely homozygous,
High levels of autozygosity associated with selfing can reduce  the effective population size of a selfing species to  less than  $1/2$ of the same outcrossing population\cite{Pollak1987,Nordborg1997}. 
Therefore, all else being equal, neutral diversity in selfing taxa should be no less than half of that observed in their outcrossing relatives.  
As selfing species often exhibit a greater than two-fold reduction in diversity, severe founding bottlenecks 
	are often presented to explain this discordance (e.g. in \rube{} \cite{Guo2009,Foxe2009}); however, alternative explanations, including the greater reach of linked selection in selfing populations have also been proposed \cite{Charlesworth1993,Cutter2003,Hedrick1980,Charlesworth2001,Baudry2001} (see below).
Such founding bottlenecks are seen as evidence supporting the idea that selfing species are often founded by a small number of individuals, 
	consistent with reproductive assurance favoring the evolution of selfing \cite{Fisher1941 ,Schoen1996}.
	%	rather than the gradual evolution of selfing favored by other advantages of selfing, such as the `automatic advantage' of fertilizing yourself as well as other plants \cite{Fisher1941 ,Schoen1996}. 

%Although both a bottleneck at the origin of  \rube{} and an `Out-of-Greece' bottleneck play an important role in shaping patterns of diversity in \rube{,}
The very low levels of diversity within \rube{} seemed initially to be consistent with this view \cite{Guo2009}.  
Indeed, we find that for a given genomic region, few founding lineages drawn from a \grand{-}like population contributed ancestry to present day \rube{}. 
However, this reduction in \emph{C. rubella}'s diversity relative to \emph{C. grandiflora}, and the observation of only one or two 
	extant % I THNIK YOU MEAN TO SAY EXTINCT, Dr Coop
	founding haplotypes in most genomic regions (as previously observed \cite{Guo2009}) is due to an extreme loss of variation subsequent to the founding of \rube{,} 
	and  does not necessarily imply an extreme founding bottleneck. 
This loss of variation is likely due to an extreme reduction in
\emph{C. rubella}'s effective population size, the potential
  causes of which we discuss shortly.

%\gc{Graham: Deleted 2nd 1/2 of sentence, mvoe the citations you want} 
%	due to other demographic consequences of the transition to selfing (e.g. population substructure and population size fluctuations \cite{Ingvarsson2002,Wright2003}), 
%	and the longer reach of linked selection with a lower effective population recombination rate \cite{Charlesworth2001}. 

The high level of drift due to this small $N_e$ confounds our ability to estimate the actual number of founding chromosomes, 
	because the genetic contribution of founders has been lost (see  \cite{Anderson2007, Leblois2007} further discussion). 
We therefore caution that low long-term effective population sizes in
	selfing plants may  erode historical signals of their founding. 
Our likelihood based inference as well as our evidence for more than two founding haplotypes in some genomic regions argues against the hypothesis that \rube{} was founded by a single plant with no subsequent secondary contact from \grand{;} however, we lack sufficient information to pinpoint the founding population size.

%\paragraph{Long term reduction in \rube{} effective population size:}
%The observations above reflect the fact that the size of the population that founded \rube{} played a minor role in shaping \rube{} diversity, as compared to drift since its founding. 
%In principal, this small population size could reflect slow population growth after the founding of \rube{.} 
%However, we do not find much evidence suggesting recent population growth within our founding lineages (although we cannot rule out very recent growth due to our sample size, see \cite{Adams2004}).
%Instead, 
The patterns of diversity that have arisen since \emph{C. rubella}'s founding are consistent with a population 
	at approximately mutation-drift equilibrium with a small long-term effective population size. 
In fact, we estimate a twenty-fold reduction in \emph{C. rubella's} effective number of chromosomes from the $\approx$ 600,000 in \grand{.}
Although the  causes of this reduced effective population size  are unclear, 
	numerous forces, including frequent oscillations in population size, linked selection, etc. may be responsible \cite{Charlesworth2001,Ingvarsson2002,Wright2003,Wright2008}, 
	and future work on the determinants of $N_e$ in selfing species will clarify this issue.

This small effective population size has led to a rapid loss of diversity since \emph{C. rubella}'s founding.
While some genomic regions maintain multiple extant founding lineages and high levels of pairwise sequence diversity, 
	if this small size persists \rube{} will quickly lose much of its genetic variation. 
For example, currently two individuals inherit the same founding haplotype for approximately $80\%$ of the genome, resulting in a profound lack of diversity.
At the current rate, it will take only another $40$ky for $95\%$ of the genome of two individuals to be homozygous for all ancestral variation. 
This would reduce genome-wide $\pi_S$ in \emph{C. rubella} to $0.0016$, severely limiting the pool of standing variation available for a response to selection.
Perhaps it is this low diversity that limits the adaptive evolution \cite{Glemin2013} of selfing species and contributes to their eventual demise \cite{Stebbins1957,Takebayashi2001,Goldberg2010}.

	\paragraph{Relaxed efficiency of purifying selection in \rube{:}}
%	The long-term reduction  in effective population size and our comparisons within and between founding haplotypes also clarifies the process of the accumulation of deleterious mutations in \rube{.}
Viewing \emph{C. rubella}'s founding haplotypes as a random draw from an ancestral \grand{}-like population, 
	we expect (and indeed observe -- \figref{pishap}{A})  comparable $\pi_N/\pi_S$ values among \emph{C. rubella}'s 
	founding haplotypes and within \grand{.}
Therefore, the founding of \emph{C. rubella} did not itself facilitate the accumulation of deleterious mutations, contrary to expectations from a model where an extreme reduction in $N_e$ at the species founding allowed deleterious mutations to markedly and suddenly increase in frequency.  
Rather, the long-term reduction in \emph{C. rubella}'s effective population size lessened the efficacy of purifying selection, 
	as is reflected by the threefold increase in $\pi_N/\pi_S$ within founding haplotypes as compared to between species, founding haplotypes, or within \grand{.} 
%Although most pairwise comparisons across the \rube{} genome reflect comparisons within a founding haplotype, 
%	$\pi_N/\pi_S$  within \rube{} is closer to interspecific $\pi_N/\pi_S$, because most of \emph{C. rubella}'s diversity ($\approx 90\%$) is among founding haplotypes. 
%Given the calculation above, if \rube{ } 
%		maintains its relatively small effective population size this ratio should become even more skewed.	
 Our view of the origin of deleterious mutations in \rube{} can reconcile two seemingly contradictory observations -- 
		that $\pi_N/\pi_S$ within selfing species is large but $d_N/d_S$ between selfers and close relatives is  unremarkable (e.g. \cite{Glemin2006}).
	The unremarkable $d_N/d_S$ between selfers and their relatives reflects the fact that since selfing species are generally young, an overwhelming portion of their divergence from outcrossing relatives is simply the sorting of ancestral variation.
	By contrast, the high $\pi_N/\pi_S$ observed within selfing species reflects the rapid homogenization of most initial variation in selfing taxa, and the weakening of purifying selection against novel non-synonymous mutations, which can make up a substantial portion of intraspecific variation while hardly contributing to interspecific divergence.

\paragraph{Future prospects:} 
With our haplotype-based approach, we provide a reasonable sketch of 
	\emph{C. rubella}'s history.
		However, numerous questions remain. 
		Future work on the population genomics of selfing will identify the cause(s) of the reduced effective population size often observed in selfing populations, highlight the role of rare introgression in the evolution of selfing, identify recent fluctuations in the size of selfing populations, and inform the geographic spread of selfing lineages. 
		While full sequence data from more individuals will further illuminate these issues, 
		our result highlight the vast information about species' origin present in population genomic data. 
		Future analyses like the one presented here will help further refine our genomic understanding of the evolutionary transition to selfing.

%	For example, with whole genome data, as opposed to transcriptomic data, we should be able to more clearly identify contiguous regions with more than two founding haplotypes. 
%	With sequence from more individuals, it may be possible to identify regions of recent introgression between \emph{Capsella} species.
%	Moreover, additional samples will provide a finer-grained view of the allele frequency spectrum, 
%		ushering in  a more complete model of population expansion and contraction over the last 50 ky in \rube{,}
%		and may illuminate genomic patterns of positive and negative selection in a recent founding selfing species.  

% You may title this section "Methods" or "Models". 
% "Models" is not a valid title for PLoS ONE authors. However, PLoS ONE
% authors may use "Analysis" 
\section*{Materials and Methods}

	\subsection*{Sequencing, alignment, and sequence quality}
		We utilized genotype data from 38 bp paired-end
                sequencing of RNA extracted from flower bud tissue of
                11 samples (6 \rube{} and 5 \grand{}).
                These reads
                  were then mapped to the \rube{}  reference genome using Tophat \cite{Trapnell2009} (v.1.3.0)
		as described previously  \cite{Wright2012} (using an inner distance between reads (-r) of 100, and minimum and maximum intron length of 40 and 1000 respectively).
	To call SNPs from the RNA data, we utilized the GATK pipeline
        on the BAM files \cite{McKenna2010,DePristo2011}.
We instituted straightforward QC steps, and treated all genotypes with coverage less than 10X, quality scores (from the GATK pipeline) less than 30, 
		and/or heterozygous sites in putatively autozygous
                regions  as missing data. 

To validate our calls we compared our genotype data to $\approx 53,000$ sites of Sanger sequencing and found very little discordance (see  \emph{Text S1}, Table S1), and nearly identical diversity measures 
		($\pi_{\text{Sanger}} = 0.156\%$,  $\pi_{\text{Transciptome}} = 0.159\%$,  
		for 72,066 and 71,645 pairwise comparisons between base pairs, respectively).
	We analyzed all loci where individual genotypes passed quality control standards allowing us to utilize sites with partially missing data, 
		a slight departure from the initial presentation of this data set \cite{Wright2012}, which only examined sites where all individuals passed QC.  
	We focus on divergence and diversity at fourfold degenerate (i.e. synonymous) and zero fold degenerate (i.e. nonsynonymous) sites to view patterns of neutral and putatively deleterious variation within and among species.

\subsection*{Identifying allozygous regions through patterns of heterozygosity}
	
	Given the high selfing rate in \rube{,} \cite{StOnge2011} the genome of a \rube{} individual is expected to be mostly autozygous. 
	However, some allozygous regions are expected in
        field-collected samples of a species with a non-zero outcrossing rate. 
	Indeed, we observe heterozygous sites in our \rube{} samples.
	Such sites could be caused by genotyping and/or alignment error, 
		\emph{de novo} mutations,  
		 or residual heterozygosity retained since a lineage's most recent outcrossing event (i.e. heterozygosity in allozygous regions).	
	Since allozygous loci will be clustered in the genome due to the limited number of generations for recombination since the most recent outcrossed ancestor, 
		while sequencing errors will be distributed relatively uniformly across the genome, 
		we utilize the distribution of heterozygous sites across the genome to separate allozygous regions from sequencing error in \rube{.}
	More specifically, we identify allozygous regions by examining the local density of heterozygous sites. 
	These regions are generally quite obvious (Figure S9A-F), so we visually identified the beginning and ends of these allozygous stretches of the genome within an individual. 

 We treat these allozygous regions of an individual's genome as missing data. 
         Reassuringly, the average heterozygosity within an individual in these allozygous regions  ($\pi_{S\text{, ind}} = 0.43\%$) closely matches the pairwise diversity between individuals ($\pi_{S} = 0.41\%$ see Figure S7). This gives us confidence that by treating these allozygous regions as missing data for an individual 
we are not biasing ourselves away from interesting genomic regions of high diversity. 
	By contrast, nearly all heterozygous sites in putatively autozygous regions should be artifacts (e.g. sequencing error, misalignment, etc.),
	 	and very few should represent \emph{de novo} mutations that have arisen since the region was last made homozygous by descent due to inbreeding.
In inferred autozygous regions on average 0.13\% of synonymous sites are heterozygous. This error rate varies across individuals (see \emph{Text S1}, Figure S8), corresponding to sequencing lane. 
We treat these heterozygous sites in allozygous regions as missing data in our population genomic analyses.

\subsection*{Identifying founding haplotypes}
	Since \emph{C. rubella} and \emph{C. grandiflora} have recently split, much  variation within each species is incompletely sorted variation inherited from  a population ancestral to both.
	In \emph{C. rubella}, this ancestry can persist for long physical distances, due to its recent founding and low effective recombination rate.
	We can therefore hope to infer the haplotypes that contributed
        to the founding of extant {\em C. rubella} diversity. %\gc{I
                                %THINK this wording side steps what
                                %you explained}. 	
	In doing so, we do not attempt to assign founding haplotypes in regions between informative data, therefore minimizing our uncertainty in founding haplotype assignment. 

One of the strengths of this approach is that even ancestrally polymorphic alleles
that are missing from our small sample of extant \emph{C. grandiflora}
diversity, but by chance are found in our \emph{C. rubella} sample,
are likely to be correctly identified as differences among founding
haplotypes, rather than contributing to difference within founding haplotypes.
This follows from the fact that such sites will often be flanked by
jointly polymorphic sites that were common in the ancestral
population, allowing us to correctly assign the status of founding haplotype sharing.

	\emph{Preliminary haplotype assignment:}
	In some genomic regions, all of our samples will carry the same founding haplotype. 
	Thus, we assign all \emph{C. rubella} samples to the same founding haplotype in long regions ($>10$ kb and $> 4$ polymorphisms in \emph{C. grandiflora}) where all \emph{C. rubella} samples 
		(with non-missing data) are identical at positions polymorphic in \emph{C. grandiflora}.
	
We next focus on pairwise comparisons in regions where polymorphisms are jointly segregating, since such variation likely represents incompletely sorted ancestral variation. 
In regions of the genome where a pair of \emph{C. rubella}  individuals have inherited the same founding haplotype, they must have identical alleles at ancestrally polymorphic sites. 
We labeled all sites polymorphic in both species as a `same site' if both individuals were homozygous for the same allele,  
	and as a `different site' if both individuals were homozygous for different alleles. 
We labeled sites as missing data if at least one of the pair did not pass QC at this site. 
We identified runs of haplotype sharing between two samples beginning with a `same site' and ending at the last `same site' before a `different site,' ignoring sites with missing data.   
When these runs of `same'  sites extended more than 1.5 kb  and consisted of at least 4 jointly polymorphic sites, we preliminarily assigned these individuals to the same founding haplotype.
 
In regions with ancestry from exactly two founding haplotypes (e.g. the left hand side of Figure1B), alternative founding haplotypes must differ at sites polymorphic in both species -- that is, with two extant founding haplotypes, differences at jointly polymorphic sites are necessary and sufficient for assigning individuals to alternate founding haplotypes. 
%We therefore preliminary assign pairs of individuals to different founding haplotypes in regions where they are first on a `different site' and ending at the last `different site' before a `same site', if these runs persisted for $>$1.5 kb  and consisted of at least 4 jointly polymorphic sites.
In regions with more than two extant founding haplotypes, differences at jointly segregating sites are sufficient but not necessary for assigning individuals to alternate founding haplotypes, because two distinct founding haplotypes could be identical at the same jointly polymorphic allele. 
We explore alternative founding haplotype labeling rules in  \emph{Text S1}, and show that our results hold under most reasonable criteria.

	\emph{Higher order haplotype assignment:}
	Building on pairwise founding haplotype assignments, we aim to identify alternative founding haplotypes across the \emph{C. rubella} genome. 
	To do so,	we broke the genome into windows of differing sizes corresponding to points in which runs of pairwise (same vs different) founding haplotype assignment begin and end across individuals.
	We then assigned  individuals to founding haplotypes in each window as follows:
	\begin{enumerate}
		\item We did not  attempt to infer the founding haplotype of an individual in a region where it was allozygous.
		\item In invariant regions, we assigned all individuals to the same founding haplotype.
		\item In all other regions, we assigned individuals with `same' and `different' founding haplotype assignments onto alternative founding haplotypes by constructing networks of haplotype sharing. 
	To do this,  
	\begin{enumerate}	
		\item We began with the first individual (this choice does not affect the algorithm, see below) and found which (if any) others where on the same founding haplotype by the above criteria, and labeled all individuals as `founding haplotype one'.
		\item We continued this process until no individuals are the same as founding haplotype one. 
		\item We then chose the first individual not assigned to founding haplotype one, and place it on founding haplotype two, finding the other individuals inferred to have inherited this founding haplotype as described above. 
		\item We continued this scheme, introducing additional founding haplotypes as necessary (i.e. repeating step 3), until all of these individuals where assigned to a founding haplotype. 
	\end{enumerate}

	\item Occasionally, we could not assign an individual to a founding haplotype in a region, and so we labeled this individual as `ambiguous'. 
	This could occur for two reasons. The first is that due to missing data, there was discordance in our founding haplotype assignment, 
		e.g. individual 1 was assigned to the same founding haplotype as individual 2 and 3, but individuals 2 and 3 were assigned to different founding haplotypes. 
	To be conservative in such cases we labeled all three (or more) individuals as `ambiguous'  
		this both minimizes uncertainty and ensures that how we assign individuals in our algorithm does not influence our results. 
The second reason for an individual to be assigned an `ambiguous' label is because pairwise assignments began and ended at the first and last different (or same) ancestrally polymorphic site, in some regions an individual was not assigned to the same or different founding haplotypes as any other samples. 
These regions could represent an individual switching rapidly between founding haplotypes
	 due to historical recombination events, or a third founding haplotype present only once in our sample.
	\end{enumerate}
At the conclusion of this algorithm every individual was assigned to a founding haplotype (or labelled as ambiguous) for every genomic window where an individual was autozygous. 
 We do not use these  ambiguous regions  when comparing within or among founding haplotypes, and we examine the possibility of regions with more than two founding haplotypes in the main text.

\subsection*{Constructing neighbor joining trees}
We used the \emph{nj} function in the R \cite{R} package ape \cite{ape} to construct neighbor-joining trees 
	(presented in \figref{pishap}{C}) from distance matrices containing subsets of our SNP data set at synonymous sites.
For the entire transcriptome (\figref{pishap}{C1}) we constructed the distance matrix  
	where each off-diagonal element was the fraction of pairwise sequence differences 
	between the pair of individuals ($i$ and $j$) at synonymous sites, 
	$\pi_{S,ij}$ where $i$ and $j$ refer to rows and columns of the distance matrix. 
%For the trees for within and among haplotypes we again used the nj package, using a different definition of the distance matrix. 
For the tree constructed within \emph{C. rubella}'s founding haplotypes (\figref{pishap}{C}2), 
	we calculated the fraction of  pairwise sequence differences between the pair of \rube{ }  individuals ($i$ and $j$) 
	where we inferred $i$ and $j$ to have inherited the same founding haplotype. 
For the tree constructed among \emph{C. rubella}'s founding haplotypes (\figref{pishap}{C}3), 
	we calculated the fraction of  pairwise sequence differences between the pair of \rube{ }  individuals ($i$ and $j$)  
	where we inferred $i$ and $j$ to have inherited different founding haplotypes. 
In both cases, entries in the distance matrix between pairs of \grand{} and \rube{,} and within \grand{} pairs where constructed by using all synonymous sites. 
We note that numerous recombination events clearly occurred during the history of these samples, and we therefore caution against interpreting this neighbor joining tree as a phylogenetic statement.

%	For trees within and among haplotypes, off-diagonal entries between \emph{C. rubella} samples were $\pi_{U,W,ij}$) and $\pi_{U,A,ij}$, respectively. 

\subsection*{Demographic inference}
	To infer the history of \rube{,} we simulated a coalescent model where at time $\tau$,  
		$N_f$ chromosomes founded a population that instantaneously grew to $N_0$ effective chromosomes (Figure 1A). 
	To avoid potential confusion with the definition of the effective population size in selfers (see \cite{Balloux2003} for recent discussion), 
	we directly used the effective number of chromosomes, 
		$N_0$, as our coalescent units, so that  the rate of coalescence of a pair of lineages equaled $1/N_0$. 
	We note that our inference of the number of founding chromosomes was inspired by two recent papers \cite{Anderson2007,Leblois2007} 
		that addressed this question using small numbers of micro-satellite and PCR amplified loci, respectively.

To infer the demographic parameters of interest ($\tau$, $N_f$, and $N_0$),  we made use of 
	the frequency with which all samples are assigned to the same founding haplotype, $\eta$,  
	and the allele frequency spectrum in these regions, $\phi$. 
In our four exchangeable individuals (three Greek and one Out-of-Greece), $\eta=.66$, and $\phi = \{.62 \text{ singletons : } .22 \text{ doubletons : } .16 \text{ tripletons}\}$.
We aimed to estimate the composite likelihood of our data given our parameters, $C(\phi, \eta| \tau, N_0, N_f)$, via coalescent simulation.
	As this likelihood depends on only $\tau/N_0$ -- the coalescent-scaled founding time, and not on $\tau$ and $N_0$ separately, 
we estimated the likelihood surface as a function of this compound parameter $C(\phi , \eta  | \tau/N_0, N_f)$. 
We then resolved these two parameters by considering nucleotide diversity within founding haplotypes (below). 

For inference, we use a composite likelihood framework.
Composite likelihoods approximate the full likelihood of
the data as the product of the likelihoods of a set of correlated observations --  ignoring their dependance.
This facilitates inference in cases where obtaining the full likelihood is
computationally prohibitive   (see  \cite{Adams2004, Gutenkunst2009, Hudson2001} for earlier
population genetic applications).  
In making this approximation, composite likelihoods make the likelihood
surface overly peaked, but do not produced a bias in the maximum
likelihood estimate (MLE) \cite{Larribe2011, Wiuf2006}.

\subsubsection*{Coalescent simulations}
We found $C(\phi , \eta  | \tau/N_0, N_f)$ by generating expectations $\phi$ and $\eta$ from 10,000 coalescent replicates across each cell in a fine-grained grid of  $\tau/N_0$ and $N_f$.
Specifically, we simulated the coalescent genealogy of four lineages in a population with $N_0$ effective chromosomes, back to time $\tau/N_0$.
	For a given simulation, our sample of four had coalesced to $n_c$ lineages ($1 \leq n_c \leq 4$) at time $\tau/N_0$. 
	With probability, $(1/N_f)^{(n_c-1)}$, all $n_c$ lineages coalesced to the same founding haplotype, at time $\tau$, 
		and with probability $1-(1/N_f)^{(n_c-1)}$ we expected more than one extant founding haplotype. 
	For each simulation, we kept track of 
		the proportion of simulations where all  samples coalesced to the same founding haplotype ($\eta_{\text{simulated}}$), 
		and  a vector of the time with $k$ lineages, $T_k$ ($2\leq k \leq 4$).

\emph{Likelihood of the allele frequency spectrum, $\phi$:} 
	We used this distribution of coalescence times to calculate the expected allele frequency spectrum within a founding haplotype, 
		$\phi$, by computing the expected number of sites with $i$ copies of a derived allele, $E[\xi_i]$, from  \cite{Griffiths1999} 
\begin{equation}
	E[\xi_i] = \frac{2 N_0\mu}{i} {{n-1} \choose i}^{-1} \text{   } \sum \limits_{k=2}^{n-i+1}   {k \choose 2}  {{n-k} \choose {i-1}} E[T_k] 
		\text{  \hspace{2cm} } 1\le i \le (n-1)
	\label{GriffithsTavare}
\end{equation}
Where $\theta$ is the population mutation rate.
We then converted  $E[\xi_i]$ into the expected proportion of polymorphic sites observed $i$ times in a sample of size $n$, $	E[\phi_i] = E[\xi_i] / (\sum E[\xi_i])  $, 
	i.e. the expected frequency spectrum conditional on all four samples inheriting the same founding haplotype.
Since $E[\phi_i]$ is independent of $N_0\mu$ this value allowed us to disentangle $N_0$ and $\tau$, below. 
The probability of an allele frequency spectrum across many unlinked sites is multinomial with probabilities given by $E[\phi_i]$ and the number of observations (i.e. the number of polymorphic sites within our four samples in regions where we inferred all to have inherited the same founding chromosome), which we used to estimate the composite likelihood of $\phi$ given the parameters, $C(\phi |  \tau/N_0, N_f)$.

\emph{Likelihood of the proportion of the genome derived from a single founding haplotype, $\eta$:}   
	The probability that all samples coalesce to the same founding haplotype is binomial with probability $\eta_\text{simulated}$, 
		 which we used to estimate the likelihood of $\eta_\text{estimated}$ given the model. 
	A difficulty with estimating the likelihood of  $\eta$ is that  there is no natural observable unit for a founding haplotype to take a product of likelihoods over. 
	We took a conservative solution to this challenge -- 
		since most regions where individuals share a founding haplotype are shorter than $1$cM, 
		and since our map covered $\sim 300$ \emph{cM}, we conservatively assumed that we observed $100$ independent founding haplotype regions.

\emph{Disentangling founding time ($\tau$) and current population size ($N_0$):}  
Using neutral diversity within the founding haplotypes used in this analysis, 
	an estimate of $\mu$ ($1.5 \times 10^{-8}$ \cite{Koch2001}, as above), 
	and estimates of $N_f$ and $\tau/N_0$, we could  estimate $N_0$ independently of $\tau$.
Via simulation, we found the expected number of generations since two lineages coalesce conditional on these lineages inheriting the same founding haplotype. 
We solved this to match the average $\pi_S/2\mu$ within haplotypes of our four exchangeable samples to obtain an estimate of $N_0$.
\section*{Acknowledgments}
We would like to thank Dan Koenig, David Haig, Molly Przeworski, Michael Turelli, Jeremiah Busch, Peter Ralph, Alisa Sedghifar, Jeremy Berg, Mike May, and Gideon Bradburd for their thoughtful comments.
%This work was made possible by a NSF Bioinformatics postdoctoral fellowship to YB, support to GC from the Sloan Fellowship in Computational and Evolutionary Molecular Biology, NSERC and the Genome Quebec and Genome Canada VEGI grant (StW), and the Swedish Research Council (TS).

% FUNDING ETC.

%\section*{References}
% The bibtex filename
\bibliography{Capsella1.bib}

\newpage
%\newpage
%%%%%%%%%%%%%%%%%%%%%%%%%%%%%%
%%%%%%%%%%%%%%%%%%%%%%%%%%%%%%
\section*{Tables}
\begin{table}[ht]
\begin{center}
\emph{Table 1}: Percent sequence variation within and between \emph{Capsella spp.}
\begin{tabular}{|c|c|ccc|}
  \hline
  & &  & \emph{C. rubella}  & \\
 & \emph{C. grandiflora} & Greek & Out-of-Greece & All \\ 
% & \emph{C. grandiflora} &  \emph{C. rubella} & Out & rube \\ 
  \hline
\emph{C. grandiflora} & 1.86 [1.83,1.93]  &  &  &   \\ 
\hline
  Greek  \emph{C. rubella} & 2.03 [2.01,2.13] &    0.40 [0.39,0.45] &  & \\ 
  Out-of-Greece  \emph{C. rubella} & 2.02 [2.00,2.09] & 0.46 [0.45,0.52] &  0.27 [0.26,0.32] & \\   
All  \emph{C. rubella} & 2.03 [2.00,2.11] 	& NA &  NA &  0.41 [0.40,0.46] \\ 
   \hline
\end{tabular}
\end{center}
\end{table}

	Table 1: {{\bf Neutral variation within and between
            \emph{Capsella} populations:}} Percent sequence
        differences at synonymous sites averaged across pairs of individuals within and between \emph{C. rubella} and \emph{C. grandiflora}. 
This matrix is symmetric and comparisons between partially overlapping
sets (e.g.\emph{C. rubella} x Greek \emph{C. rubella}) are noted as 'NA'. Redundant cells above the main diagonal are intentionally left blank. 
%%%%%%%%%%%%%%%%%%%%%%%%%%%%%%
%%%%%%%%%%%%%%%%%%%%%%%%%%%%%%
\newpage
\section*{Figures}
%Fig1

\newpage

%Fig2

%Fig3
\newpage
%Fig4

%Fig5

\clearpage
\newpage

\section*{Text S1}
	Supplementary text: {{\bf Additional details about the data and analyses presented above:}}
	$1)$ Data summary without reference to haplotype.
	$2)$ The $f_3$ test shows no signal of recent introgression. 
	$3)$ Founding haplotype sharing by physical distance. 
	$4)$ Validation of RNA-Seq genotypes by Sanger sequencing. 
	$5)$	Robustness of results to thresholds for calling haplotype.  
	$6)$	A summary of diversity within and among founding haplotypes. 
	$7)$  $\pi_S$  and major haplotype frequency of the across all chromosomes. 
	$8)$ Putatively allozygous regions. 
	$9)$ Third haplotype candidate regions.
\newpage

%%%%%%%
%SUPP MAT
%%%%%%%

\section*{Data summary without reference to haplotype:}
We present basic data summaries without reference to our haplotype-based method in  Figure S1, as discussed in the main text.

\section*{The $f_3$ test shows no signal of recent introgression}
	To further investigate the possibility of recent introgression in Greece, we made use of the $f_3$ statistic \cite{Patterson2012,Reich2009}. 
	This test compares the covariance of allele frequencies in three populations to identify a signal of gene flow \cite{Patterson2012,Reich2009}. 
	$f_3(c;a,b)$  is formally defined as $E[(c'-a')(c'-b')]$, where the prime denotes allele frequencies in putatively admixed population, $c$ and putative source populations, $b$ and $c$, 
		and can only be negative if population c is a mixture of populations closely related to populations $a$ and $b$.
	We found $f_3\text{(Greek; {\em C. grandiflora}, Out-of-Greece})$ to be significantly greater than zero (0.35  [0.32,0.38] ), and therefore lack evidence for recent admixture between Greek \emph{C. rubella} and \emph{C. grandiflora}.
%	Therefore, although we cannot fully rule out rare or old introgression events, 
%		it is clear that introgression has not been common enough to be a dominant influence on the frequency of common alleles in \emph{C. rubella}.
\section*{Founding haplotype sharing by physical distance}
In the main text, we presented summaries of distances and proportions of founding haplotype sharing. 
In those analyses, we measured distance on a genetic map generated in a \emph{C. rubella} x \emph{C. grandiflora} interspecific cross.
In Figure S2  we show that  our summaries of founding haplotype sharing qualitatively hold when measuring in physical, rather than genetic distance. 
Across all pairwise comparisons, we observe a slightly higher proportion of the genome inheriting the same founding haplotype when comparing within Greece, as compared to between Greek and Out-of-Greece samples, while we see the most sharing of founding haplotypes in comparisons between Out-of-Greece samples.

	\section*{Validation}
	We compared our genotype calls to $\approx 53$ Kb of Sager sequencing (see Table S1A for sequenced regions / effort) to empirically  investigate the error rate of our data.
	On the whole, there was little discordance between sequences (2 miscalls 52306 bp, an error rate more than an order of magnitude lower than $\pi_S$ within founding haplotypes), and $\pi$ for both data types was remarkably similar. 
	In Table S1B  we present a summary of our comparisons between RNA-Seq and Sanger sequencing.
	The two miscalls both occurred in our Argentinean sample. 
	Both of these samples were run on the same and both of which exhibited higher levels of heterozygosity in putatively autozygous regions than the other individuals (Figure S8, below), constant with a lane effect on sequencing quality.  
	Below we discuss  how this potential lane effect on error rate could influence inference.

	\subsection*{Potential influence of errors on inference}
	The potentially higher error rate in our Argentinian and Algerian samples has relatively little influence on our major conclusions.
	These two samples were not used in our demographic models, and the (still low) error rate is  too low to have a substantial influence on genome-wide diversity measures. 
	Additionally, since sequencing errors are likely overwhelmingly singletons, they are unlikely to influence our haplotype labeling, 
		which makes use of common variants shared across species. 
	
	However, such sequencing errors could influence two summaries of diversity within founding haplotypes in the Out-of-Greece samples: 
	\begin{enumerate}
		\item  {\bf{Overestimate of Out-of-Greece growth:}} We observed an excess of singletons in Out-of-Greek samples residing  the same founding haplotype, suggesting recent growth and/or significant population structure Out-of-Greece (main text). However, this results is also consistent with sequencing error, which generates singletons, and therefore some of this signal may be due to sequencing error. Thus, although we have clear evidence for an Out-of-Greece history of \emph{C. rubella}, the rate of population growth and/or population structure outside of Greece is unclear.
		\item {\bf{Overestimate of $\pi_N /  \pi_S$ within haplotypes in Out-of-Greece samples:}}  Within founding haplotypes, diversity at synonymous relative to non synonymous sites ($\pi_N /  \pi_S$) increases with the number of Out-of-Greece samples. 
	Since sequencing error is expected to target sites without respect to their degeneracy, while purifying selection is expected to eliminate deleterious mutations, sequencing error can increase 	$\pi_N /  \pi_S$ and therefore may contribute to the high $\pi_N /  \pi_S$ observed outside of Greece.
	\end{enumerate}
	In summary, the pattern of potential sequencing error may change some details of the diasporan history \emph{C. rubella} but does not strongly influence our major findings regarding the history of \emph{C. rubella}.

\section*{Robustness of results to haplotype calling cutoffs}

In the \emph{METHODS},  we describe our algorithm for haplotype assignment. 
This algorithm requires us to prescribe threshold values for the number of consecutive SNPs and the distance in base pairs required to assign individuals to the same founding haplotype 
	(i.e. in our pairwise assignment we insist that two individuals are identical at sites polymorphic in both species for $X_{\text{SNP}}$ over at least $Y_{\text{BP}}$). 
We then combined information across individuals to create `higher order assignments,' where we assigned all individuals to the same founding haplotype when there was no joint polymorphism for ten kb and five  SNPs.
Here we show that our major conclusions are robust to these cutoffs by demonstrating that inference is consistent across a diversity of  pairwise combinations of $X_{\text{SNP}} = \{2, 4, 10\}$ and $Y_{\text{BP}}  = \{10, 10^3, 10^4, 10^5\}$).

While all major results hold across all parameters investigated, some of the details change.
Below, we discuss how these change what influence our parameters have on some informative summary statistics, and how these results alter the interpretation of our findings.

\subsection*{Haplotype assignment and haplotype sharing:}

As  the criteria for assigning individuals to the same or different founding haplotype became stricter (e.g. $X_{\text{SNP}}$ and/or $Y_{\text{BP}}$ increased),  
	proportionately less of the genome provided clean haplotype calls (exactly one or two founding haplotypes), while more of the genome yielded ambiguous haplotype calls and/or was inferred to contain more than two founding haplotypes (Table S2). 
These results are consistent with expectations,  increasing the stringency necessary to assign individuals to founding haplotypes left us with fewer regions where individuals can be assigned to founding haplotypes.  
This expected effect also influences the portion of samples assigned to the same or different founding haplotypes across geographic comparisons  (Figure S3, compare to Figure 2A in the main text).

	\subsection*{Summary statistics}
	In Figures S4 and S5, we present basic summaries of variation within and among founding haplotypes across haplotype labeling cutoffs.

	We present all three-way allele frequency spectra within haplotypes averaged across geographic comparisons in Figure S4.
	Note that, although some results change slightly with haplotype calling rules, results are relatively stable and consistently separated from both the standard neutral expectations and the allele frequency spectrum without reference to founding haplotype. 
	
	We present $\pi_S$ and $\pi_N/\pi_S$ within and among founding haplotypes in Figure S5. Although results are relatively consistent across parameters, there are a few noteworthy trends. 
	\begin{enumerate}
		\item  {\bf{Same haplotype:}} Insisting  on  strict criteria to assign chromosomes to the same founding haplotype results in these regions being very recently diverged (e.g. Figure 3B), 
		and, as expected decreases $\pi_S$ within founding haplotypes (S5A). 
		This recency appears to also result in less time for putatively deleterious mutations to be removed from the population, increasing $\pi_N/\pi_S$ within founding haplotypes (S5A).
		\item  {\bf{Different haplotype:}} Perhaps counterintuitively, increasing the length for which two samples must differ at sites polymorphic in both species also decreases $\pi_S$ and $\pi_N/\pi_S$ between  founding haplotypes (S5B). This result could be due to a slight enrichment of regions in which all samples reside on the same founding  haplotype too short to be caught by our \emph{ad-hoc} rules. However, diversity among  founding haplotypes is orthogonal to our major questions and inferences, and therefore this result does not influence our major claims in any way.
	\end{enumerate}

	\subsection*{Inference}
	Above, we showed that most of our summary statistics do not change, or change slightly and predictably across founding haplotype calling thresholds.
	Here, we review our three main results concerning the history or \emph{C. rubella} gleaned from our coalescent model and investigate how founding haplotype assignment cutoffs influence these conclusions 
	\begin{enumerate}
		\item	 {\bf{No need to postulate an extreme bottleneck:}} 
			 In the main text, we showed that while we could not completely rule out an `extreme' founding of \emph{C. rubella}, we had little evidence supporting this hypothesis. 
			 We arrive at a similar conclusion for most founding haplotype calling rules (Figure S6A); 
			 	however, when only exceptionally long regions can be assigned to founding haplotypes, our model begins to favor an extreme founding event.
			 This result  is expected -- as we limit the regions assigned to founding haplotypes these regions will seem young and long and will trace their ancestry to few founders.
			 Since even under these standards, a large number of founders is still likely, and since this extreme method of haplotype calling is expected to generate this bias,  
			 	we find little compelling evidence for an `extreme' founding event.
		\item	 {\bf{Reduced long term  effective population size:}} We infer a very small effective population size ($N_0 = \{25,000 - 40,000\}$) across all haplotype labeling cutoffs. 
		We find a decrease in the inferred $N_0$ with the stringency  required to assign  samples to founding haplotypes. 
		Two potential factors likely generate this pattern
			\begin{enumerate}
			\item Shared ancestry across long distances suggests recent common ancestry and hence less time for mutations to accumulate 
				(a result observed in this data, but not presented), decreasing $\pi_S$.
			\item Lower stringency may accidentally place samples on the same haplotype, artificially increasing estimates of $\pi_S$.
			\end{enumerate}
		\item	 {\bf{\emph{C. rubella} originated $\approx 50$ kya:}} Estimates of the date of origin of \emph{C. rubella} vary slightly across founding haplotype calling threshold for reasons similar to those listed above. 
		The 95\% confidence intervals are partially overlapping for every date estimate provided. Note that these confidence intervals are larger than those provided in the main text because here we do not constrain our initial population size to be the MLE.
		We note that the variation in our estimates induced by our haplotype labeling rules pales in comparison to our uncertainty in the mutation rate.   
		If we replace the estimate of $\mu$/gen of $1.5 \times 10^{-8}$, with an alternative estimate of $7 \times 10^{-9}$  we double our estimated split time (see main text).
	\end{enumerate}

\section*{A summary of diversity within and among founding haplotypes}
Figures 3A and 3B , we display $\pi_S$ and $\pi_N/\pi_S$ within and among \emph{C. rubella's} founding haplotypes. 
In table S3, we present these values and the associated bootstrapped confidence intervals.

\section*{$\pi_S$ and major haplotype  frequency of the across all chromosomes:}
	In the main text we present the relationship between nucleotide diversity and haplotype frequency for chromosome seven (Figure 5).
	We present similar results across all chromosomes in 10 kb windows with a 2 kb slide, below (Figure S7).
	As in Figure 5, synonymous diversity is in purple, 
		the inferred major haplotype frequency is in orange, and red points  putatively containing more than two founding haplotypes.
 	Also, like Figure 5, nucleotide diversity increases as major founding haplotype frequency decreases.

\section*{Allozygous regions}

	We present individual heterozygosity at synonymous sites   in putatively allozygous and autozygous genomic regions in Figure S8. 
	Diversity in allozygous regions closely matches diversity between individuals, as expected if regions that we infer to be allozygous are correctly identified. 
	While individual heterozygosity is clearly higher in allozygous than in autozygous regions, we still observe heterozygous genotypes in putatively autozygous regions. 
	We treat these heterozygous sites in putatively autozygous regions as missing data since hey likely represent sequencing errors, 
		and we point out that they are overrepresented and are predominantly singletons in the Algerian and Argentinian samples (consistent with the potential lane effect on error rate, described above). 

	We label allozygous regions by eye (Figures S9A-F) for each individual -- 
		 inferring a region to be allozygous when the slope of the cumulative number of heterozygous sites on physical position is relatively large. 	
	We exclude these putatively allozygous regions from all haplotype-based analyses.  
	Sites heterozygous in all \emph{C. rubella} samples are censured in \emph{C. grandiflora} and \emph{C. rubella} analyses, as they likely represent common misalignments.

\section*{Third haplotype candidate regions}
We display nine regions likely to contain more than three founding haplotypes in Figure S10.   
In each panel we present $\pi_S$ between all combinations of three individuals inferred to have inherited alternative founding haplotypes (across 10 kb windows each overlapping by 2 kb). 
When $\pi_S$ between each individual in the trio is high (i.e. near the level of $d_U$ -- the dashed horizontal line) it is likely that the three individuals have inherited distinct founding haplotypes. 
We also display $\pi_S$ within a founding haplotype in grey -- the small values of these grey lines argues against the possibility that these regions are poorly aligned. 
We note that this non-random sample of our 172 candidate regions was chosen to argue that some of these regions are likely correct, 
	and it is therefore that \emph{C. rubella} originated from a single founder without subsequent introgression. 
We also note that since $\pi_S$ between some samples is very low (grey lines), these regions are not easily dismissed as likely alignment errors.

\clearpage
\section{Supplementary Figures}
%S1
	\begin{figure}[hh!!]
	\begin{center}	
	\includegraphics[width = 5.5 in]{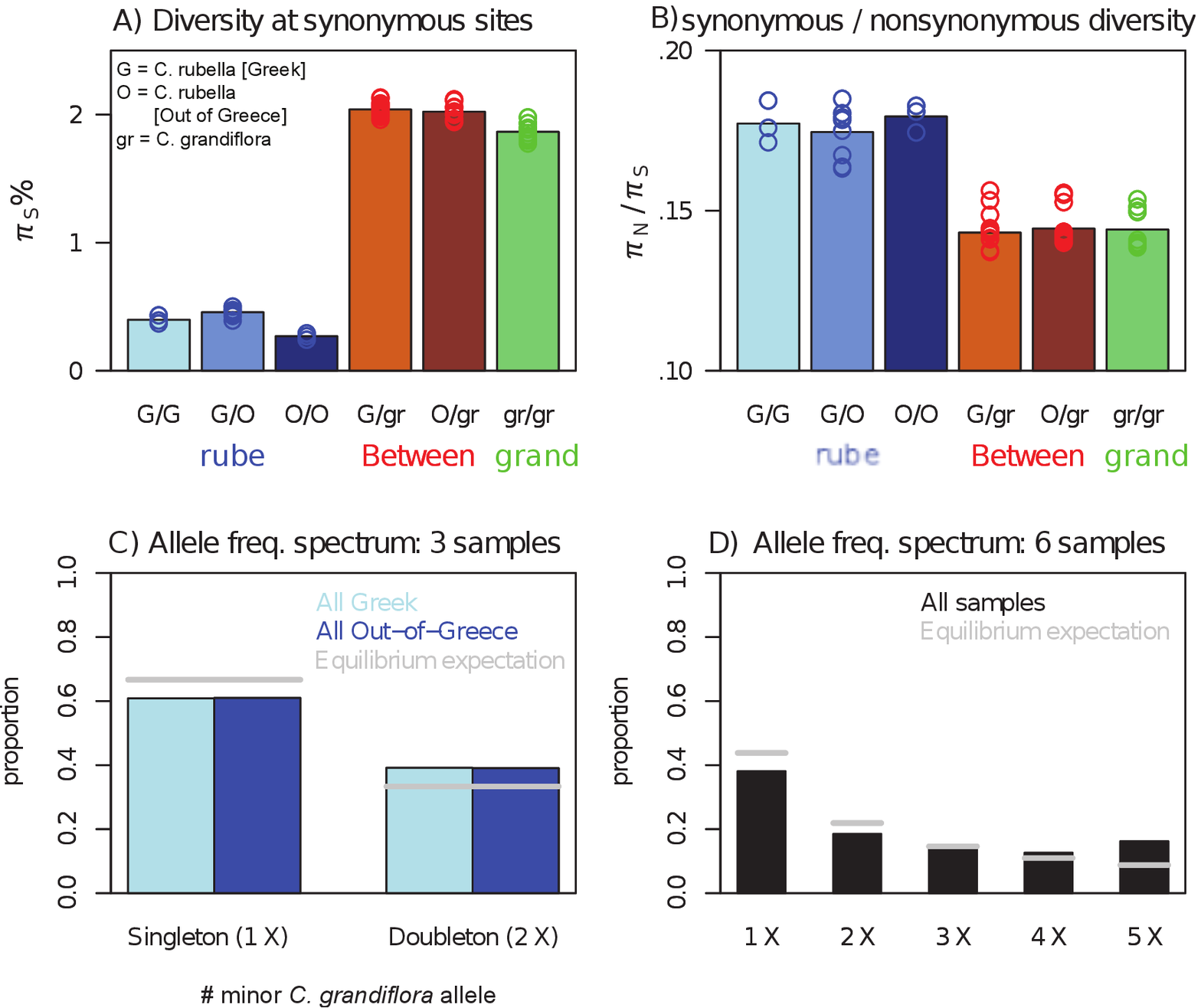}  
	\end{center}	
		Figure S1  : {{\bf{Patterns of variation in \emph{Capsella}:}} 
			$A$ and $B$) Mean number of pairwise differences at fourfold degenerate sites ($A$), and ratio of diversity at synonymous to non-synonymous sites ($B$) within \emph{C. rubella}, \emph{C. grandiflora}, and between species. 
			The type of individuals compared is displayed on the x-axis, where G = Greek \emph{C. rubell}a, O = Out-of- Greece \emph{C. rubella}, and gr = \emph{C. grandiflora}. 
			Bars represent means and points represent single pairwise comparisons. 
			$C$ and $D$) The frequency spectrum of synonymous sites segregating in \emph{C. rubella} (polarized by the common \emph{C. grandiflora} allele), within geographic comparisons ($C$) and across all samples ($D$).}	
		\label{nohap}  
	\end{figure}		
\newpage

%S2
		\begin{figure}	
			\begin{center}	
				\includegraphics[width =6 in]{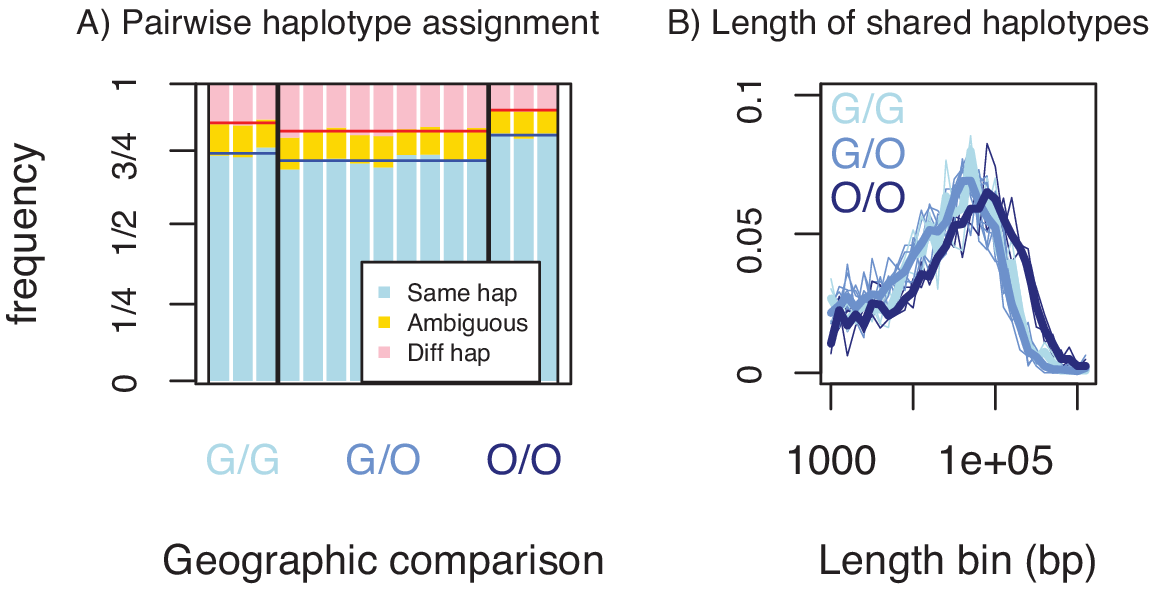}
			\end{center}	
			Figure S2:  {{\bf Patterns of haplotype sharing in \rube{:}}   $A)$ The proportion of the genome for which two individuals have inhertied the same or different founding haplotypes, or for which haplotype calls are ambiguous (e.g missing genotype makes haplotype calling problematic), by physical distance.	
			$B)$ The proportion of regions of founding haplotype sharing of length X, by physical distance. }
			\label{PHYS}
		\end{figure}	
\newpage

%S3

		\begin{figure}	
		\begin{center}	
				\includegraphics[width =4 in]{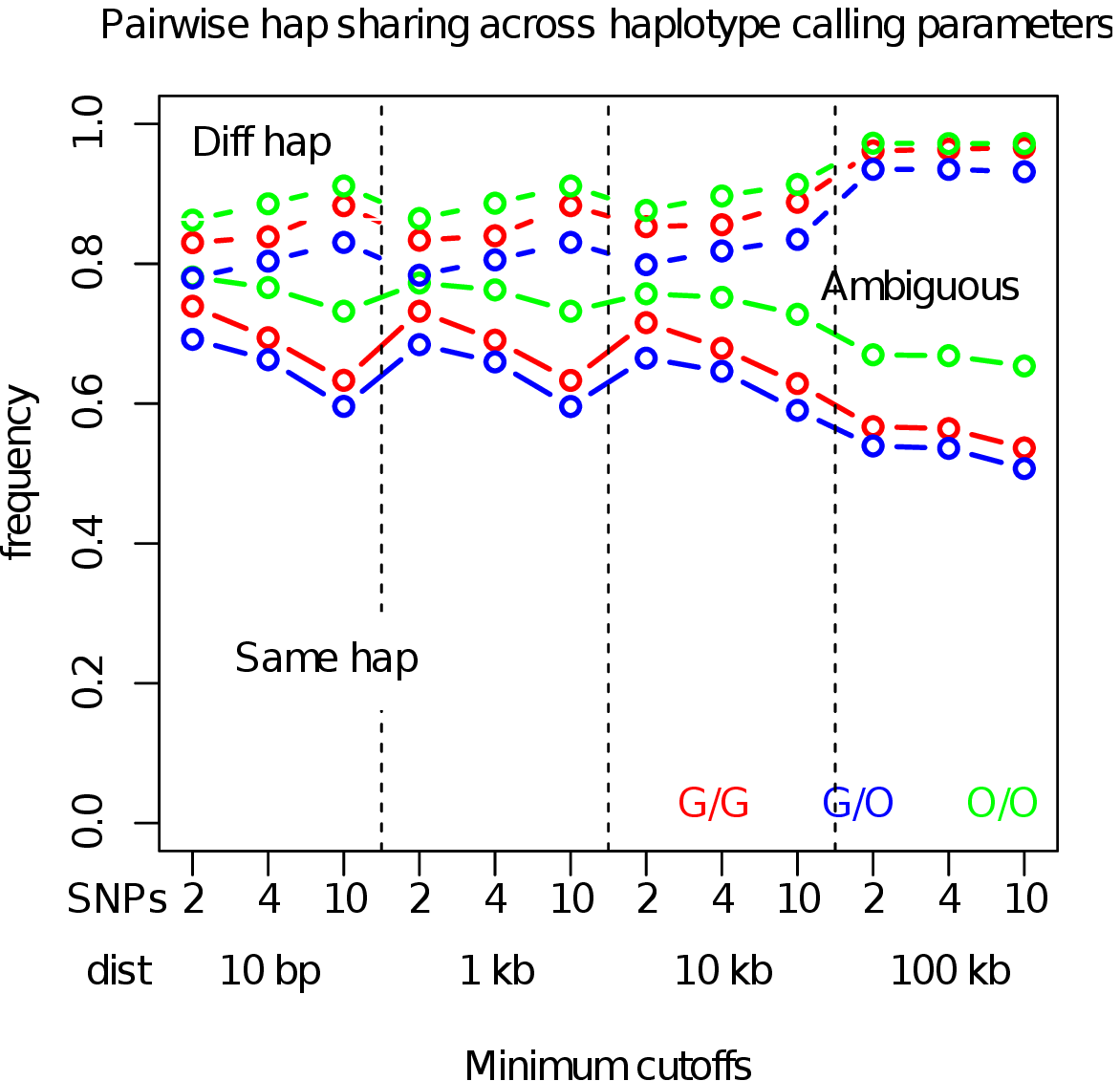}
		\end{center}	
			Figure S3: {{\bf Patterns of pairwise haplotype sharing across haplotype calling thresholds:}} 
			As haplotype calling criteria become stricter, fewer pairwise comparisons can be assigned to the same (proportion below lines) or different (proportion above dashed lines) founding haplotypes, and more become ambiguous (compare to Figure 2A in the main text).
		\label{prop}
	\end{figure}	
	\newpage

	%S4	

		\begin{figure}	
	\begin{center}	
				\includegraphics[width =4 in]{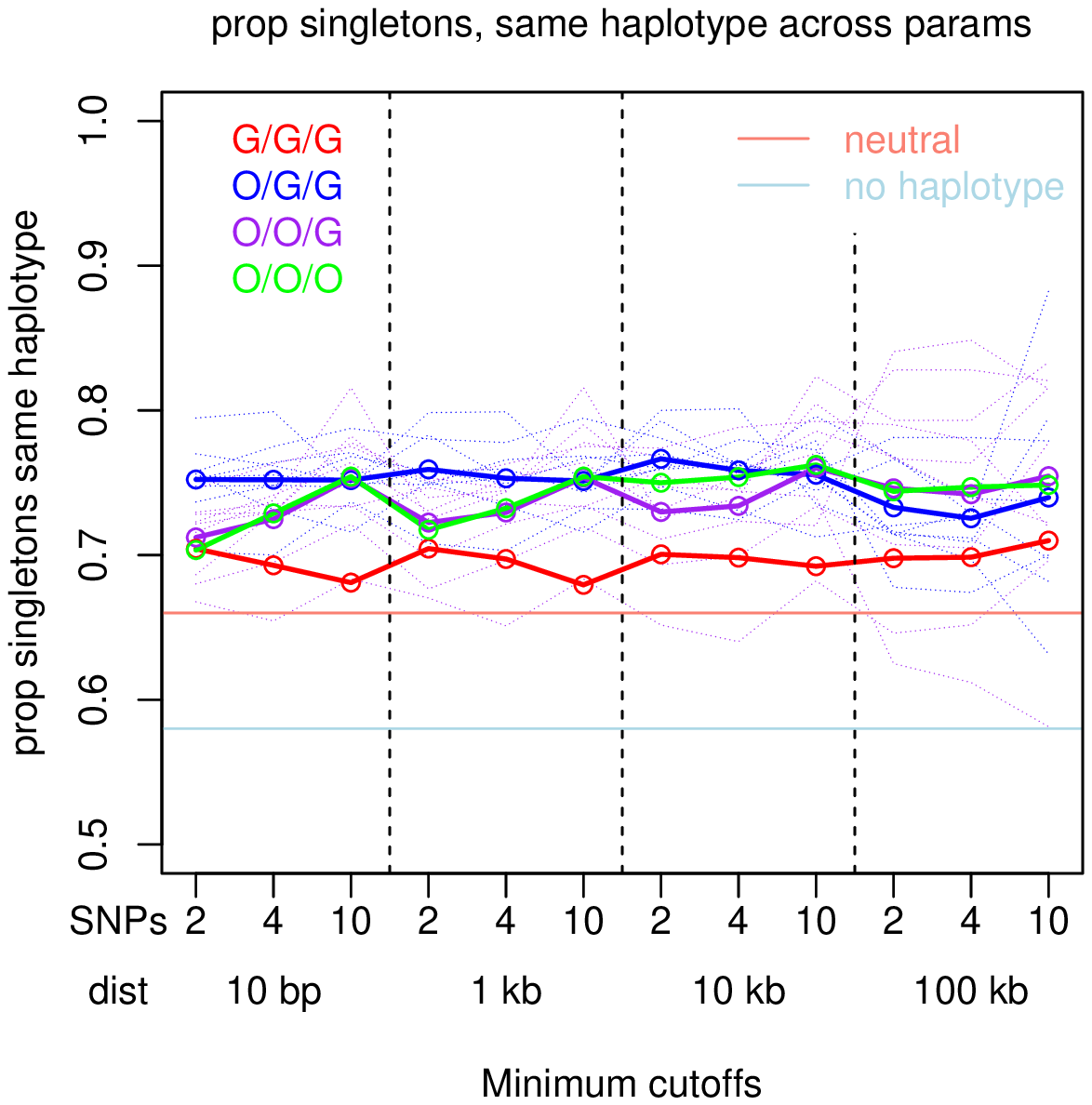}
	\end{center}	
			Figure S4: {{\bf The allele frequency spectrum (proportion of singletons in three samples) across values for haplotype calling thresholds:}} 
			Plotted separately for different types of trios -- three Greek (red), three Out-of-Greece (green), two Greek and one out of Greece (blue), or two Out-of-Greece, and one Greek (purple) samples. 
			Faint lines represent individual frequency spectra for comparisons within a trio. 
 			Thick lines represent mean values for a geographic comparison.  
			Note that the observed fluctuations are much smaller than biologically informed differences (e.g. with and without accounting for haplotype labels, or all samples within vs. some samples outside of Greece, departures from expectations under the standard neutral model).
			\label{AFS}
		\end{figure}	
		\newpage
%S5

		\begin{figure}	
		\begin{center}	
				\includegraphics[width = 4 in]{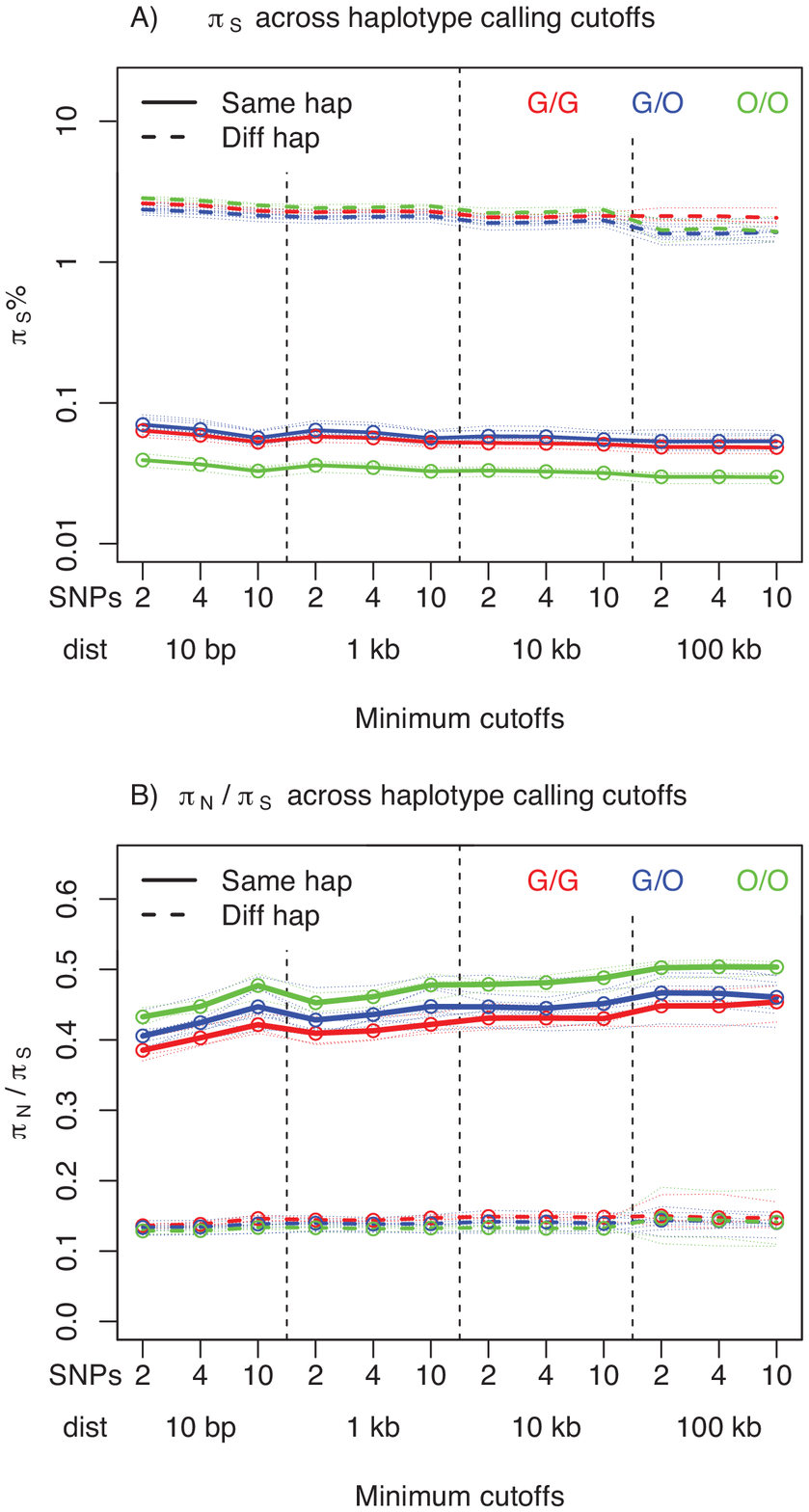}
		\end{center}	
			Figure S5: {{\bf $\pi_S$ (A) and $\pi_N/\pi_S$ (B) within and among founding haplotypes across threshold values for haplotype assignment:}} 
			Colored by geographic comparison. Lines and dashes represent comparisons within and among founding haplotypes, respectively.
			\label{PI}
		\end{figure}

%S6

		\begin{figure}	
				\includegraphics[width = 4 in]{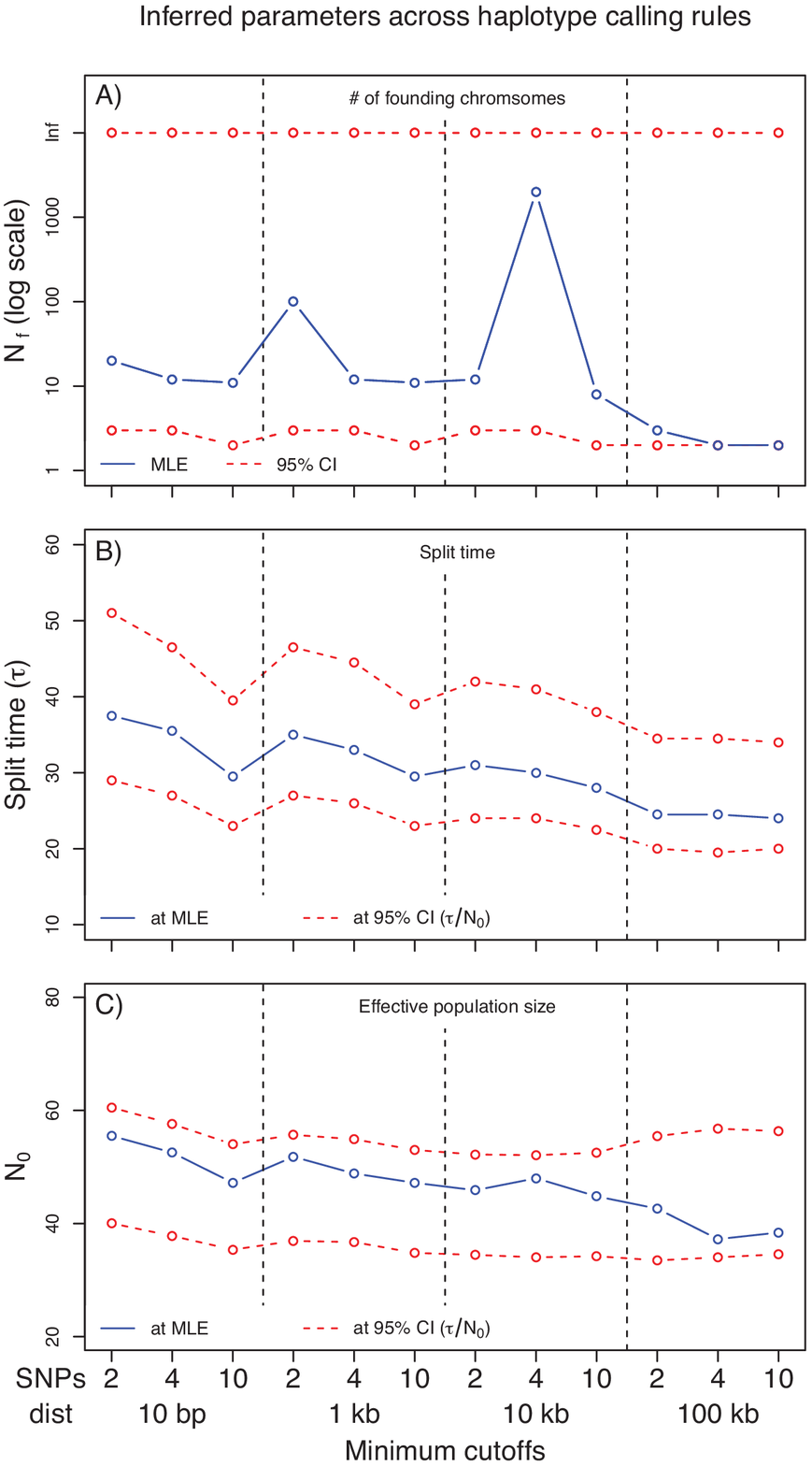}
			\begin{center}
			Figure S6: {{\bf Demographic inference across values for haplotype calling thresholds:}} 
			\end{center}
			The 95\% confidence intervals and MLEs are displayed in red and blue, respectively.  
				\emph{A)} The inferred number of founding chromosomes, $N_f$. 
				\emph{B)} The inferred split time, $\tau$. 
				\emph{C)} The inferred effective population size, $N_0$. 
%			Note that our results are largely unaffected by our choice of haplotype calling parameters, and that the in MLE are well within every 95\% confidence interval.
		\label{INFER}
		\end{figure}	
\newpage
\clearpage
%S 7
	\begin{figure}	
			%\includegraphics{figs/Chr5678.eps}
%	\hyperref[http://tinyurl.com/pu3nus5]{''http://tinyurl.com/pu3nus5'}
	Figure S7: {\bf{Haplotypic diversity and nucleotide diversity across the \emph{C. rubella genome}:}}  
	Nucleotide diversity at synonymous sites is in purple, and the inferred major haplotype frequency is in orange, while  red points are below regions putatively containing more than two founding haplotypes. Each data point represents a 10 kb window with a 2 kb slide.  
	Each of the eight panels represents a different chromosome. 	
	[link: \yb{ \url{<http://tinyurl.com/pu3nus5>} }] 
	\end{figure}
\newpage
\clearpage
% S8

	\begin{figure}	
	\begin{center}
	\includegraphics{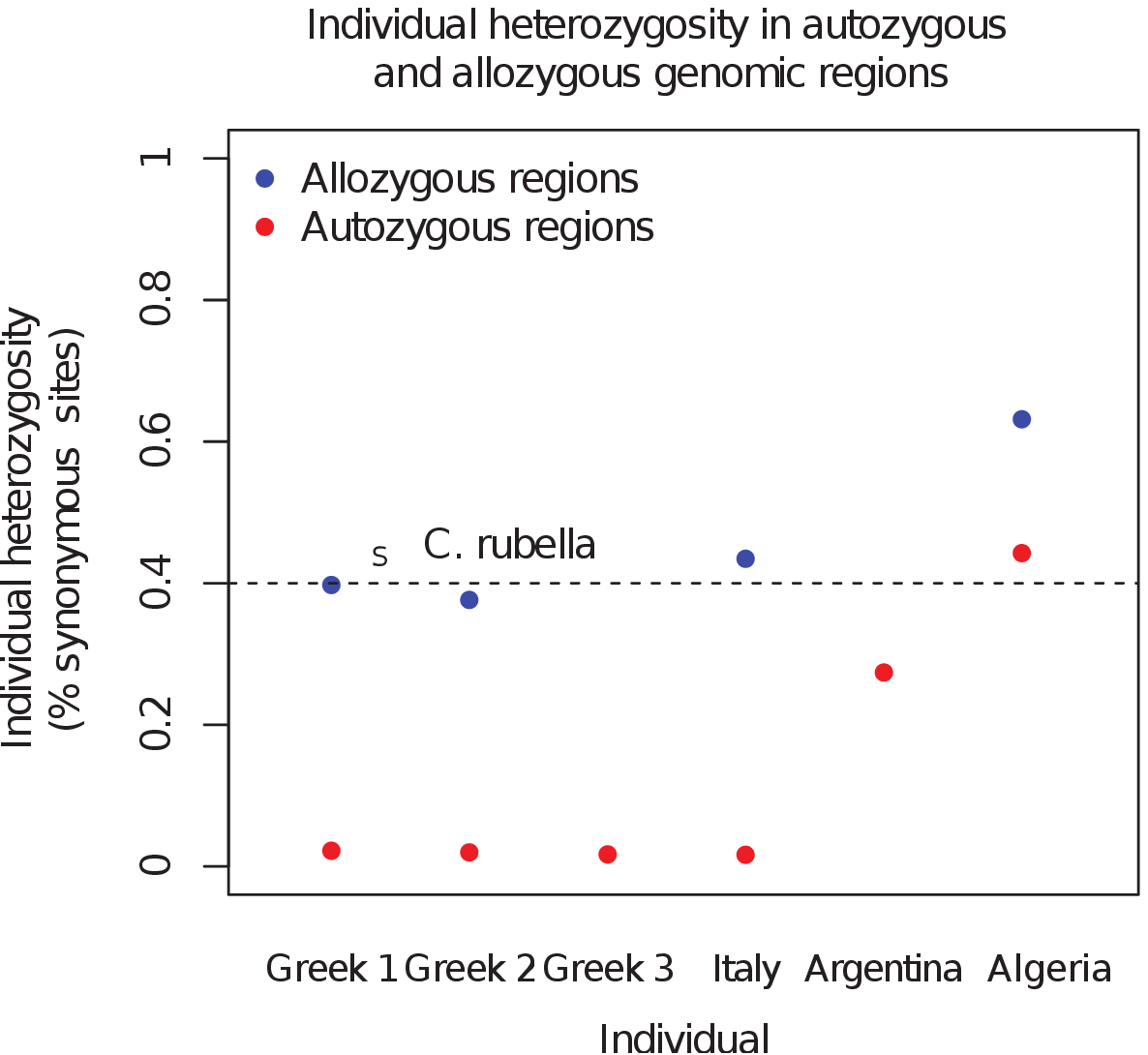}  	
	\end{center}
			Figure S8: {{\bf Genome-wide individual heterozygosity in \emph{C. rubella}:}} 
			We separately display individual heterozygosity at synonymous sites in regions inferred to be 
				allozygous (blue) or autozygous (red) for each \emph{C. rubella} individual (noted on the x-axis).   
			The dotted line represents pairwise sequence diversity between \emph{C. rubella} samples at synonymous sites for reference. 
		%	Note that these values in autozygous regions are generally very low, reflecting the low rate of sequencing error. 
			%However, two  Out-of-Greece samples (run on the same lane) had high levels of observed heterozygosity, consistent with some sequencing errors.
	%		In our analyses in the main text, heterozygous sites where treated as missing data in autozygous regions, and were randomly resolved in allozygous regions (\emph{METHODS}), but these allozygous regions where not considered in our haplotype-based inference. 
		%	Assuming these errors occur with equal likely regardless of the actual genotype, these errors will not influence our results. 
	\end{figure}		
\newpage
\clearpage
%S9
	\begin{figure}	
Figure S9: {{\bf Genome-wide individual heterozygosity in \emph{C. rubella}:}} 
	We label allozygous regions by eye (Letters A-F represent individuals, and numbers 1-8 represent chromosomes --  see figure titles). 
	Blue and red points display sites heterozygous in \emph{C. rubella} or both species, respectively, while singleton sites are presented in grey. 
	Dotted lines separate regions inferred to be autozygous and allozygous.
	The cumulative number of heterozygous genotypes is plotted on the y-axis, and the physical position is displayed on the x-axis. 
 	We infer a region to be allozygous when this slope is relatively large.
	[link: \yb{ \url{<http://tinyurl.com/qdatut7>} }] 
	\end{figure}	
\newpage
\clearpage

%S10
	\begin{figure}	
	\begin{center}
	\includegraphics{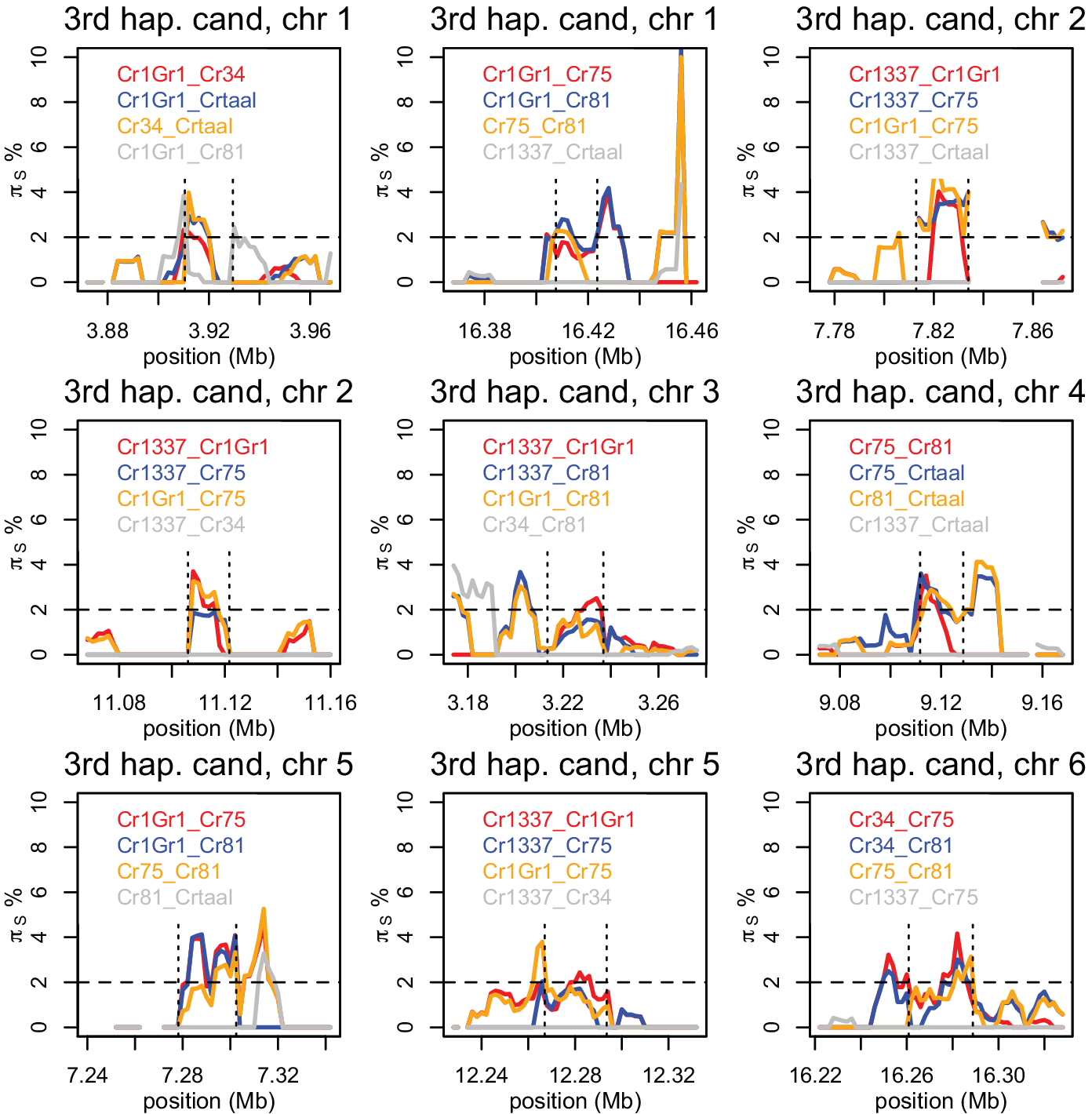}  	
	\end{center}
	Figure S10: {{\bf Evidence for more than two founding haplotypes in \emph{C. rubella}:}} 
	Nucleotide diversity ($\pi_{S}$) within \emph{C. rubella} trios in genomic regions putatively containing more than two founding haplotypes. 
	Each pairwise comparison between individuals inferred to have inherited alternative founding haplotypes is labelled in red, blue, or orange. 
	Grey lines illustrate pairwise comparisons between individuals inferred to have inherited the same founding haplotype. 
	Dashed lines mark  $\pi_S =.02$ -- the diversity expected between haplotypes on the basis of our estimates  from regions with exactly two founding haplotypes.
	Dotted lines mark the borders of genomic regions in which we infer more than two founding haplotypes.  
		Each of the nine panels in Figure S10 comprises a different region putatively containing three founding haplotypes 
		(note that the chromosome, genomic position, and trio changes in each panel).
	\end{figure}
\clearpage
\newpage
\section*{Supplementary Tables}
\begin{table}[ht]
    \begin{center}
\emph{Table S1A}: Sanger sequencing efforts.
    \begin{tabular}{lllll}
      \hline
    chrom & start & stop & orientation & inds \\   \hline
        1 & 1121520 & 1122379 & revcomp & Cr1Gr1, Cr75, Cr81, Crtaal \\ 
          1 & 2189244 & 2189862 & revcomp & Cr1Gr1, Cr75, Cr81, Crtaal \\ 
          1 & 3187217 & 3187782 & revcomp & Cr75, Cr81, Crtaal \\ 
          1 & 9252546 & 9253264 & forward & Cr1337, Cr1Gr1, Cr81, Crtaal \\ 
          1 & 9791410 & 9792193 & revcomp & Cr1Gr1, Cr75, Cr81, Crtaal \\ 
          1 & 12972008 & 12972691 & forward & Cr1Gr1, Cr81, Crtaal \\ 
          1 & 16978816 & 16979446 & forward & Cr1Gr1, Crtaal \\ 
          2 & 7158433 & 7159032 & revcomp & Cr1Gr1, Cr75, Cr81, Crtaal \\ 
          2 & 9111928 & 9112538 & revcomp & Cr75, Crtaal \\ 
          2 & 11501659 & 11502346 & revcomp & Cr1Gr1, Cr75, Cr81, Crtaal \\ 
          2 & 13499343 & 13499957 & forward & Cr81 \\ 
          3 & 4380447 & 4381090 & revcomp & Cr81 \\ 
          3 & 12783435 & 12784119 & revcomp & Cr1Gr1, Cr75, Cr81, Crtaal \\ 
          7 & 2781533 & 2782194 & forward & Cr1Gr1, Cr75, Cr81, Crtaal \\ 
          7 & 3433282 & 3434023 & revcomp & Cr1Gr1, Cr81, Crtaal \\ 
          7 & 3458572 & 3459198 & revcomp & Cr1Gr1, Cr75, Cr81, Crtaal \\ 
          7 & 4820394 & 4821059 & forward & Cr1337, Cr1Gr1, Cr81, Crtaal \\ 
          7 & 6079504 & 6080090 & forward & Cr1337, Cr1Gr1, Cr75, Cr81, Crtaal \\ 
           7 & 9123830 & 9124515 & forward & Cr1Gr1, Cr75, Cr81 \\ 
           7 & 9214127 & 9214749 & forward & Cr1Gr1, Cr75, Cr81, Crtaal \\ 
          8 & 2578408 & 2579001 & revcomp & Cr1Gr1, Cr75, Cr81, Crtaal \\ 
          8 & 6671378 & 6671912 & revcomp & Cr1Gr1, Cr75, Cr81, Crtaal \\ 
          8 & 7784867 & 7785350 & forward & Cr1Gr1, Crtaal, Cr1337, Cr75 \\ 
          8 & 9670696 & 9671200 & forward & Cr1Gr1, Cr75, Cr81, Crtaal \\ 
       \hline
    \end{tabular}
\end{center}
\end{table}

\begin{table}[ht]
\begin{center}
\emph{Table S1B}: Comparison of genotype calls across technology (Sanger/RNA-Seq).
\begin{tabular}{rrrrrrrrr}
  \hline
Sample     & 0/0     & 2/2     & 0/NA     & NA/0     & 2/NA     & 0/2        &1/1     & NA/NA \\  \hline
Cr1Gr1     & 12808     & 25         & 6        & 92         & 0         & 0        &0    & 50 \\ 
  Cr75     & 10611     & 19         & 5         & 53         & 0         & 0        &0    & 28 \\ 
  Cr81     & 13494     & 6         & 37         & 104     & 0         & 0        &1    & 50 \\ 
  Crtaal     & 12904     & 8         & 369     & 139     & 2        & 2        &0    & 50 \\ 
  Cr1337     & 2429     & 0         & 12         & 0         & 0         & 0        &0    & 13 \\ 
   \hline
\end{tabular}
\end{center}
\end{table}
	Table S1: {{\bf Validation of \emph{C. rubella} genotype calls by Sanger sequencing:}}
	$A)$ Genomic regions targeted for Sanger sequencing. The chromosome, genomic location, orientation of each ORF, and the individuals sequenced  is noted in each column. 
	$B)$ Concordance between Sanger and RNA-seq genotypes calls, split by individual. Numbers refer to the  number of inferred non-reference alleles (e.g. 0 = homozygous for the reference, 1= heterozygous, 2 = homozygous non-reference), and NAs mark missing data or data that did not pass QC. Sanger genotypes are presented before, and RNA-Seq genotypes are presented after the `/'. 
	Note the minimal discordance between genotype calls by technology. 
\newpage

\begin{table}[ht]
\emph{Table S2:} Summary of haplotype information across haplotype calling cutoffs \newline
\begin{tabular}{ccccc}
  \hline
 & $X_{\text{SNP}}$ & $Y_{\text{BP}}$ & unambiguous and  & ambiguous  and/or \\
  & &   &      $< 3 $ haplotypes &       $>2$ haplotypes  \\  \hline
   & 2     & 10 bp   & 0.77 &0.23\\
   & 4     & 10 bp  & 0.70 & 0.30 \\ 
   & 10     & 10 bp  & 0.58 & 0.42 \\ 
   & 2     & 1 kb   & 0.76 & 0.24 \\ 
   & 4     & 1 kb   & 0.70 & 0.30 \\ 
   & 10     & 1 kb & 0.58 & 0.42 \\ 
   & 2     & 10 kb   & 0.72 & 0.28 \\ 
   & 4     & 10 kb   & 0.67 & 0.33 \\ 
  & 10     & 10 kb  & 0.57 & 0.43 \\ 
   & 2     & 100 kb  & 0.54 & 0.46 \\ 
   & 4     & 100 kb   & 0.53 & 0.47 \\ 
   & 10     & 100 kb   & 0.49 & 0.51 \\ 
   \hline
\end{tabular}
\end{table}
	Table S2: {{\bf Influence of founding haplotype stringency on founding haplotype assignment:}}
	The proportion of the genome in which we infer less than two founding haplotypes and no ambiguity in founding haplotype assignment as we change the X consecutive SNPS spanning Y base pairs required for founding haplotype assignment. 
\newline \newline \newline	
\newpage
\begin{table}[ht]
\begin{center}
{\em Table S3)}  Variation  within and  among  \emph{C. rubella} haplotypes:
\begin{tabular}{lccccc}
  \hline
 &all & G/G & G/O & O/O    \\  \hline
  $100 \times \pi_S\%$ within haps &  5.29 [5.11-5.40] 	& 5.55  [5.24-5.84]  & 6.05 [5.78-6.24] & 3.48 [3.20-3.66]  \\   
  $\pi_S\%$ among haps  &2.01 [1.98-2.04]  & 2.02 [1.97-2.05]	& 2.00 [1.96-2.03]   & 2.03 [2.00-2.07] \\ 
  $\pi_N/\pi_S$ within haps & .438 [.425,.449] 	& .416 [.392,.441]  &.439 [.422,.458] & .461 [.426,.498] \\ 
  $\pi_N/\pi_S$ among haps &  .140 [0.139,0.142] 	& .139 [.137,.141]    & .139 [.139,.140]  & .137 [.135,.139]  \\   
   \hline
\end{tabular}
\end{center}
\end{table}
	Table S3: {{\bf  Summary of synonymous and non-synonymous variation within and among \emph{C. rubella's} founding haplotypes:}}
	All = comparison between all \emph{C. rubella} samples, 
	G/G = comparison between two \emph{C. rubella} samples from Greece,
  	G/O = comparison betweenGreek and Out-of-Greek  \emph{C. rubella} samples, 
	O/O = comparison between two \emph{C. rubella} samples from Out-of-Greece. \\
\end{document}